# The Calendars of India

By

Vinod K. Mishra, Ph.D.

# Preface.

While growing up in India, I was always intrigued by the New Year's Day. It was a holiday and a long awaited one by office workers and village people alike. But the month name of January did not occur in the traditional month names used by my village. No one could answer me satisfactorily why we did not have our own New Year and why it was not celebrated. This fact bothered me for some time, but then I got used to it.

Later during my adult years I came across many different kinds of New Years. Also I found out that there were not one but many New Years celebrated across the length and breadth of India. This is very unusual because among all the ancient civilizations (like, Chinese, Jewish, Christian, etc.) Indians alone do not have a unique calendar. Depending on one's point of view, this may be seen as a celebration of India's diversity or an indictment of her disunity down through the ages.

I wanted to understand this diversity and so I started writing a book to explain it to myself. Now a Calendar is a product of a culture's scientific achievements and theological assumptions. I have tried to look at both the angles for understanding the *panchā˙nga* (traditional Indian Calendar). The background information about astronomy and some history has been included to put matters in perspective.

I would like to acknowledge the help I got from many books, articles, and of course from the Internet.

# 1. Introduction

India is a land of diversity with many religions, languages and regional cultures. This even gets carried to the calendars that govern peoples' social and religious lives. If one asks many Indians, when is the Indian New Year's day, it is very easy to get many answers. The western Christian calendar (henceforth called the Common Era calendar) is the only one being followed by all Indians. Compared to many ancient cultures like Chinese, Jewish, Muslim and others, India may be unique in not having developed a single unique calendar for all of her people. Instead, we have too many calendars to choose from and all of them are based on the ancient science of *jyotiṣa*.

The word *jyotiṣa* in Sanskrit is equivalent to the two modern subjects of astronomy and astrology. It was considered to be an integral part of the ancient Vedic curriculum. The *jyotiṣa* or Science of Light is one of the six ancillary Vedic sciences, other five being Rituals (*kalpa*), Phonetics (*śikṣā*), Etymology (*nirukta*), Grammar (*vyākaraṇa*) and Prosody (*chhanda*). They were known collectively as Limbs of Vedas (*vedāṅga*) with *jyotiṣa* being known as the Eyes of *veda-puruṣa* (Vedic scriptures envisioned as a human being).

The importance of *jyotiṣa* arose from the need to understand and predict various celestial events, and fix the dates of ritual and religious significance. Later it developed to include such branches of Mathematics as Arithmetic,

Geometry, Algebra, etc. The ancient Indian scholars of *jyotiṣa* subsequently made many significant contributions to the fields of Astronomy and Mathematics.

The *jyotiṣa* comprised of both Astronomy and Astrology, as they were never considered separate in India as it happened in the West. One should emphasize the fact that Astrology has always been an integral part of Hindu Dharma (a better synonym for Hinduism). One of its theological assumptions is the correspondence between the outer world of objects and the inner world of consciousness. This is expressed by the saying "*yathā piṇḍe tathā brahmāṇḍe*", or whatever is inside the human body is out there in the outside universe. This led to the hypothesis of a deep correlation between the moving astronomical bodies of the sky and the past, present, and future lives of the newborn. The ideas of *karma* (action) and *punarjanma* (rebirth) then imply that the planetary positions are indicators of the quality of the past lives and do not have any causative functions. So in India, astrology was not justified by postulating the influence of planets on humans through physical forces (like gravitation and electromagnetism) as was attempted in the West.

In the present work, we will try to understand the astronomical and historical origins of the main calendars prevalent in India. The main part of the book consists of five chapters. The first two (ch. 2 and 3) provide the astronomical understanding and the rest (ch. 3, 4, and 5) describe the *panchāṅga*. Some of the relevant information has been provided in the Appendixes.

The material presented does not claim originality and has been compiled from many sources like books, magazines and Internet. I acknowledge my debt to all of them. Some of them have been mentioned by name but many have not for which I offer apologies.

## The transliteration scheme

The rules for the Roman script transliteration of Sanskrit words are given below. This system is the standard scheme followed by academic scholars. The Sanskrit or *devanāgarī* alphabet is given as:

Vowels:

*a ā i ī u ū ṛ*

*e ai o ou aṃ aḥ*

Consonants:

*ka kha ga gha ṅ*

*cha chha ja jha ña*

*ṭa ṭha ḍa ḍha ṇa*

*ta tha da dha na*

*pa pha ba bha ma*

*ya ra la va śa ṣa ṣa ha*

A few widely understood English versions of Sanskrit words have been retained.

# 2. Basic Astronomy behind the Calendars

The starting point for calendars is the definition of the basic unit of time measurement. This unit is called a **second** and is defined as 9, 192, 631, 770 periods of vibration of Cesium-133 atomic hyperfine transition. The larger units of calendrical times are days, months, and years.

## *2.1 Different Kinds of Days*

Many different definitions of a day are given in the table below.

**Day Definitions**

| Name | Definition | Present Value |
|---|---|---|
| Solar Mean Day (MD) | Mean value of the duration of solar days taken over a year on Earth. | 24h 00m 00s (by definition). |
| Siderial Day (SD) | Earth's rotational period as seen in the fixed star frame (FSF). | 23.93447 h =86164.1 s = 23 h 56 mm 4.1 s |
| *tithi* | The time interval corresponding to the angular separation between Sun and Moon by $12^0$. | Length of *tithi* is variable. |
| Mean *tithi* | Mean value of the *tithi* = 1/30 of the Siderial Lunar Month. | 0.98435 d = 23 h 37 m 28 s |

(A useful relation between day and seconds: 1 s =0.000 011 6 d, and 0.0086 s = 0.000 000 1 d)

The Earth’s rotation is responsible for the day and night. If we can view the solar system from above the ecliptic plane, then the Earth will be seen rotating as well revolving around the Sun in a counter clockwise manner.

## *2.2 Different Kinds of Months*

Month (*māsa* in Sanskrit) derives its significance and definition from the periodic movement of the moon. The lunar motion is quite complex so there are many different possible definitions of the month.

There are a number of different 'periods' for the Moon leading to many different kinds of months. These are described below.

### 2.2.1 Synodic Month

Looking at the Moon from the Earth, a very clear cycle is visible: the cycle from New Moon, through crescent, half and gibbous Moons, to the Full Moon, and back again. The diagram below shows the beginning and end of the cycle. It should be noted that the new moon occurs when Earth, Sun and Moon are in a line with Moon in the middle.

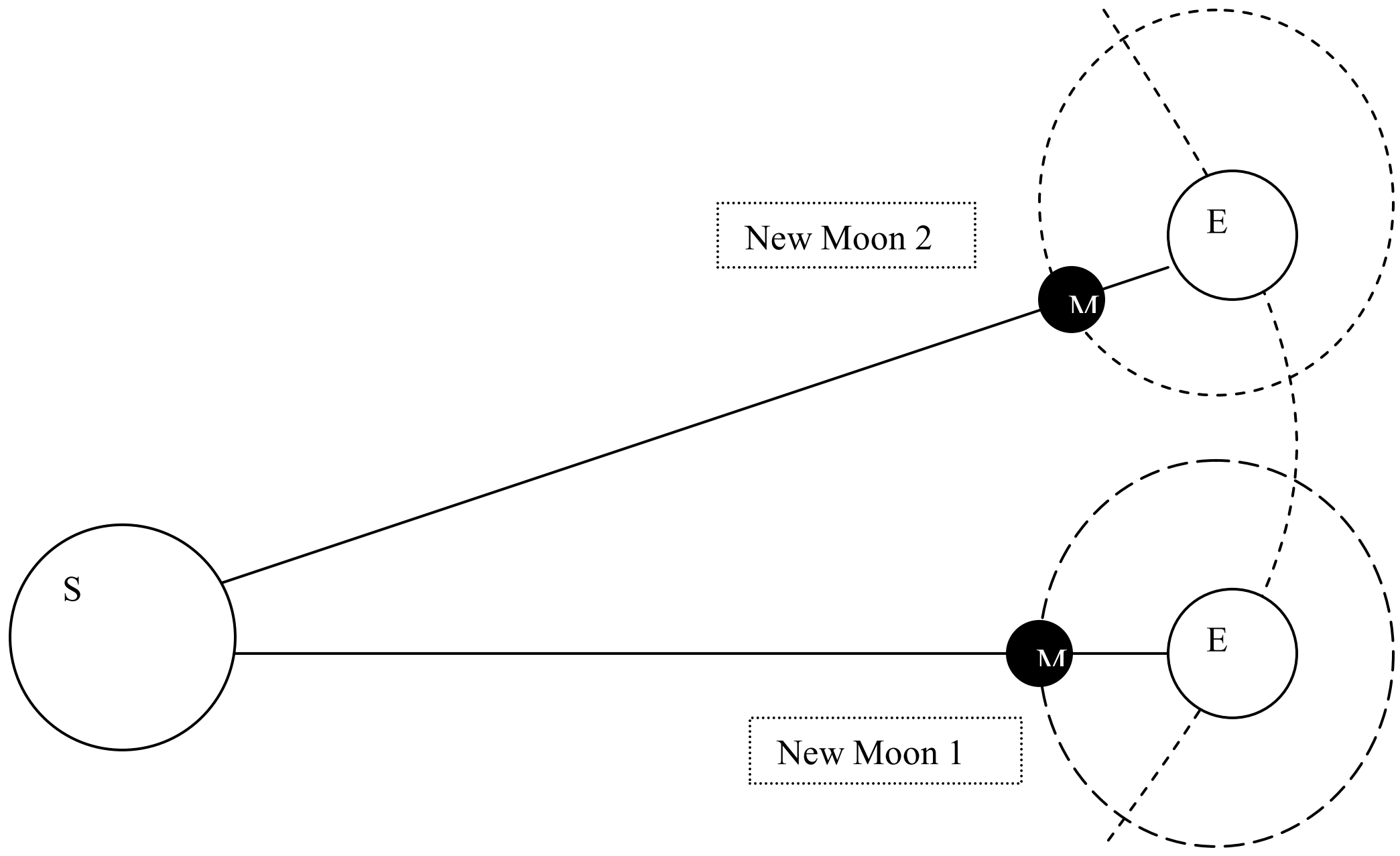

The length of this cycle is called a 'synodic' month and it is 29.530589 days on average. It is the period between lunar phases as measured from the Earth. Because of the perturbations of the orbits of the Earth and Moon, the actual time between lunations may range from about 29.27 to about 29.83 days.

**Lunar Day**: The length of a day on the Moon is the time it takes for the Sun, as seen from an observer standing on the Moon, to go from overhead to overhead. Again, this is not the rotational period of the Moon, because the Moon has moved round the Sun during that period; so, a Lunar day is the same as the time it takes for the Moon to go from full to full: ie. one Synodic Month, or 29.530589 days.

### 2.2.2 Sidereal Month

The Moon's *orbital period* is the time it takes for the Moon to complete one orbit around the Earth; i.e. to go from "due West" of the Earth, say, in the diagram below, once around the Earth, and back to "due West" again. This period is known as a **Sidereal Month**, and is 27.321661 days.

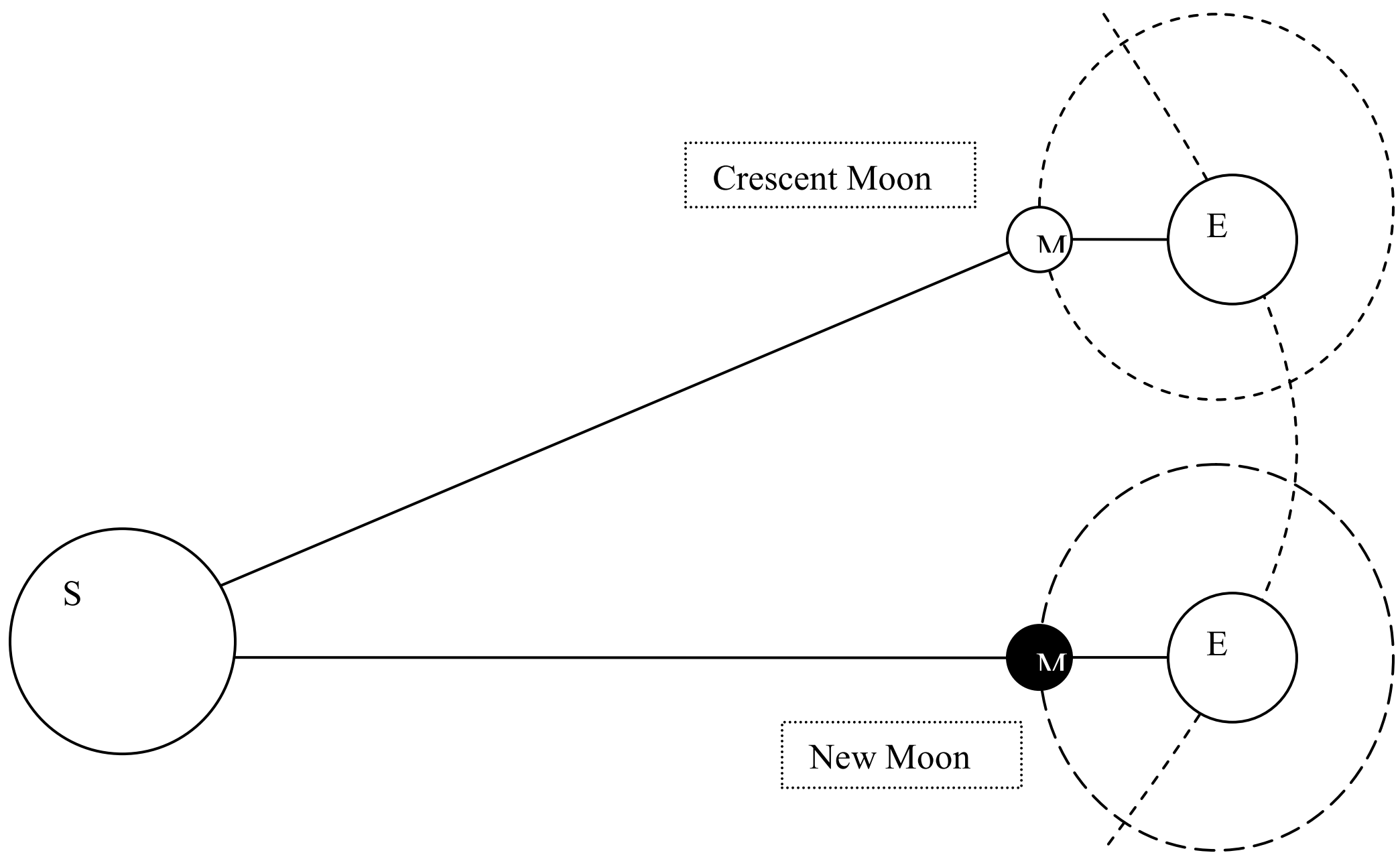

At a first glance this period may look to be the same as the synodic month but it is not so. Now in the time it takes the Moon to orbit the Earth, the Earth itself has moved, so that the Moon isn't between the Earth and the Sun when the siderial month is up -- and, therefore, isn't a New Moon. So this type of month is somewhat different from what is expected. It is not possible to measure this period with the naked eye observation.

Also, the rotational period of the Moon or the time it takes for the Moon to complete one rotation on its axis is exactly the same as the sidereal month. It is so because the Moon's rotation is synchronized, by tidal forces, with its orbit around the Earth. This is why the Moon always keeps the same face turned towards us (give or take a few degrees of variation, caused by the Moon's orbital inclination). It wasn't until the Soviet spacecraft Luna 3 photographed it in 1959 that we got our first view of the other side (sometimes referred to, inaccurately, as the "dark side").

### 2.2.3 Anomalistic Month

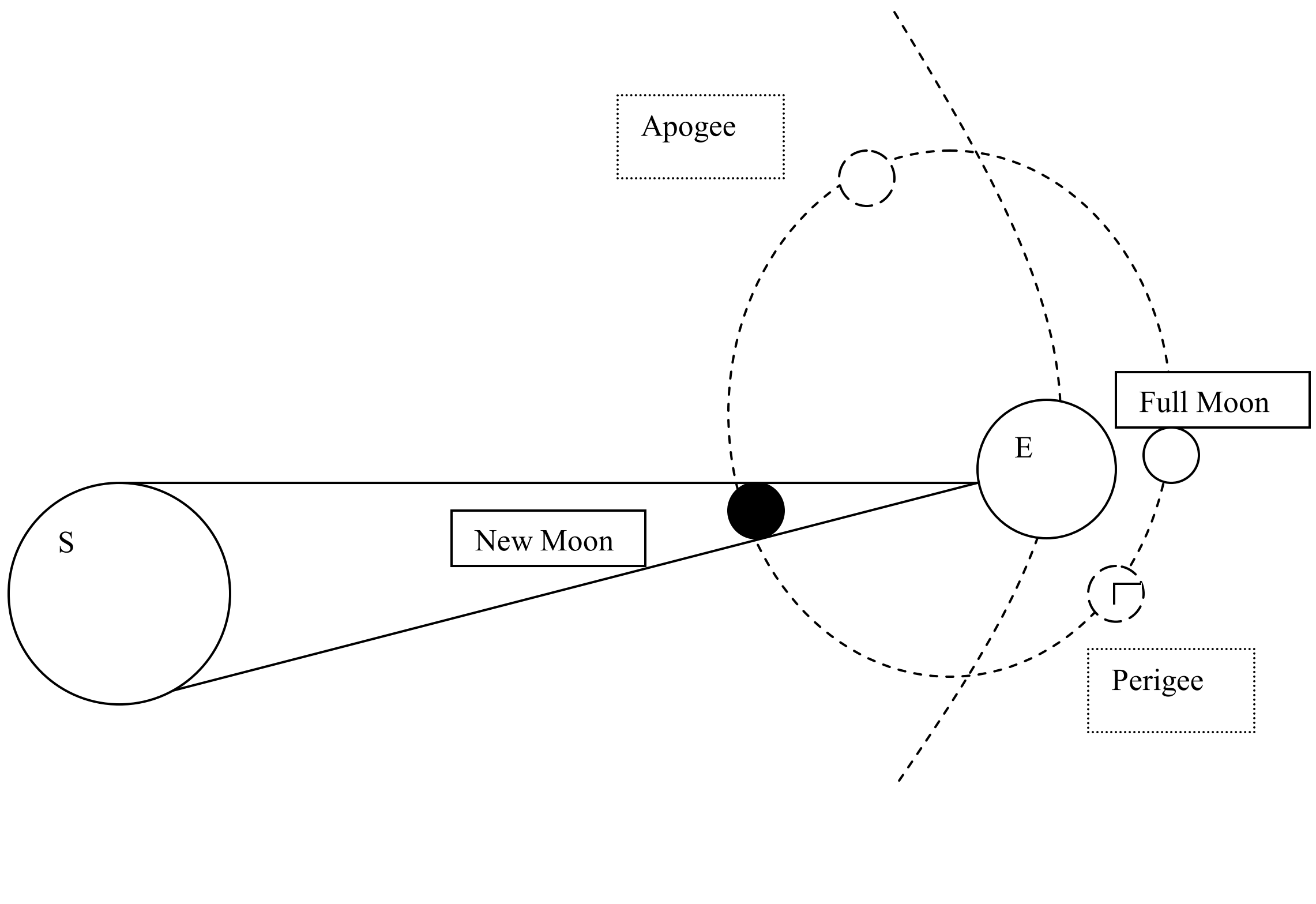

Anomalistic Month is the time taken for the Moon to go from lunar perigee to perigee.

The Moon's orbit around the Earth is an ellipse with the Moon at one focus. Apogee and Perigee are two points on this orbit that are the farthest and the closest points from the Earth respectively.

There is one more motion i.e. the regression of the Moon's orbit that one has to take into account. The longer dimension of the lunar orbit or the major axis of the ellipse (hence the apogee and perigee) circles around the Earth every 9 years. Thus, the time it takes the Moon to travel from apogee to perigee and back again is slightly longer than its orbital period or the sidereal month. This period of time is called an **Anomalistic Month**, and is 27.554549 days.

The Anomalistic Month is interesting for solar eclipses, because the "size" of solar eclipse that we see -- and the type of eclipse, whether it is partial, total, annular, or hybrid -- depend on the distance from the Earth to the Moon during the eclipse, which depends on the point during the Anomalistic Month when the eclipse occurs. This also affects the duration and the appearance of the lunar eclipses.

### 2.2.4 Draconic Month

The Moon's orbit is tilted to the plane of the ecliptic; the two points at which the orbit intersects the ecliptic are known as **nodes**. In order to get directly between the Earth and the Sun, the Moon has to be in the plane of the ecliptic; and therefore has to be at one of the two nodes. So the time the Moon takes to travel from a node, around its orbit, and back again, is of great interest.

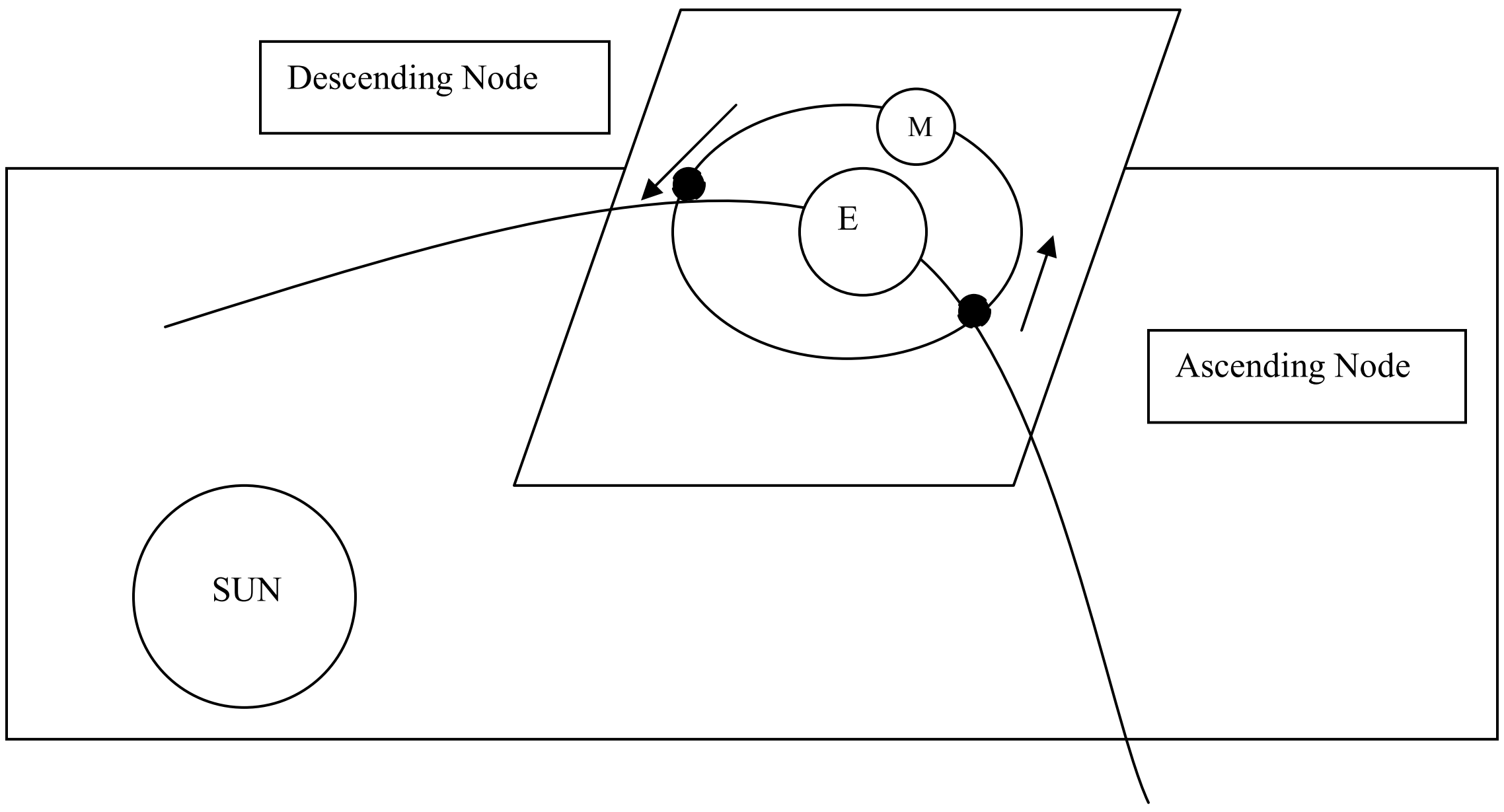

Once again, this period is not quite the same as a single orbit, due to precession of the Moon's orbit; the Moon's orbital plane, and hence the nodes, rotate backwards around the Earth about once every 18.6 years. The time taken for the Moon to return to a node is therefore shorter than its orbital period; this time is known as a **Draconic Month**, and is 27.212220 days. This is sometimes called the **Nodal Month**.

Incidentally, the term "draconic" refers to a mythological dragon that supposedly lives in the nodes and regularly eats the Sun or Moon during an eclipse. In Hindu mythology they are known as *rāhu* and *ketu* and are depicted as demons (*rākṣasa*s). Astronomers use the same words for the two nodes, which are treated as quasi-planets (*chhāyā graha*).

#### 2.2.5 Tropical Month

Just for completeness, one more kind is the **Tropical Month**, which is the period from one lunar equinox to the next where the lunar orbit crosses the Earth's equator, and is 27.321582 days.

#### 2.2.6 Other Lunar Periodicities

Motion of the Moon as seen from Earth is very complex because it is caused by so many variables in the Sun-Earth-Moon system. In addition to the values given earlier, there are some more periodicities in lunar motion, which are useful in eclipse predictions. Some of them are:

(i) The 1-year displacement period of the lunar libration longitude (Pole Nodding).

(ii) The 6-year period of the lunar libration latitude.

(iii) The 8.85-year rotation period of the Moon's perigee.

(iv) The 18.61-year Saros Cycle which is the time interval between identical lunar eclipses as seen from the same geographic location on the Earth (see Appendix E). This is also the lunar nutation period.

## *2.3 Different Kinds of Years*

The year is a larger unit of time and its definition uses the motion of the Earth or the Moon around the Sun.

The most common definition of the year is based on the revolution of the Earth around the Sun and is therefore called a `Solar Year'. However, it is possible to define the beginning and the end of one revolution in many ways and this leads to several kinds of solar years. The years so defined differ in length because of the precession of Earth's rotation and the tumbling of the Earth orbit.

Solar years have the disadvantage of not being easily observable. Many years of observations are required to fix them with any significant degree of accuracy. The use of Solar Year is thus a sign of astronomical maturity for a given civilization. On the other hand, Lunar Years were adopted first by most of the earlier cultures because the phases of the Moon -- and the first visibility after the new moon in particular -- are very easy and quick to observe.

As a consequence of Kepler's laws, the Earth moves faster in its orbit when it is near perihelion (the point in Earth's orbit closest to the Sun).

On solstices, the projection of the Earth's rotation axis on the ecliptic plane directly points to the Sun. On equinoxes, the radial line from the Sun to the Earth makes a perpendicular to the Earth's axis.

The tropical year defined by the period between equinoxes is not the same as that defined by the period between the solstices. There are two reasons for this: precession and the variability in the Earth's orbital speed depending on its position in the orbit.

### 2.3.1 Lunar Year

Before defining a lunar year, we have to define a synodic month. A synodic month is the interval from one new moon to the next as observed from the Earth, and lasts 29.5306 days. The first calendars defined a lunar year, usually consisting of 12 synodic months. The two choices have been adopted as the beginning of the synodic month, either from the new moon or from the day after full moon. In India both choices have been made among the regional almanacs.

Since for practical reasons a month should contain an integer number of days, most calendars alternated betweens months of 29 and 30 days respectively. A year made out of six months of each type has 354 days (29 x 6 + 30 x 6 = 354) and is thus too short by 0.3672 days as compared with a true lunar year. Lunar calendars have to insert one leap day about every third year to keep in step with the moon phases. On top of that, a pure or uncompensated lunar calendar is not synchronous with the seasons.

### 2.3.2 Tropical Year

A tropical year is defined as the interval between two successive passages of the Mean Sun through the mean vernal equinox and lasts 365.242199 days UT.

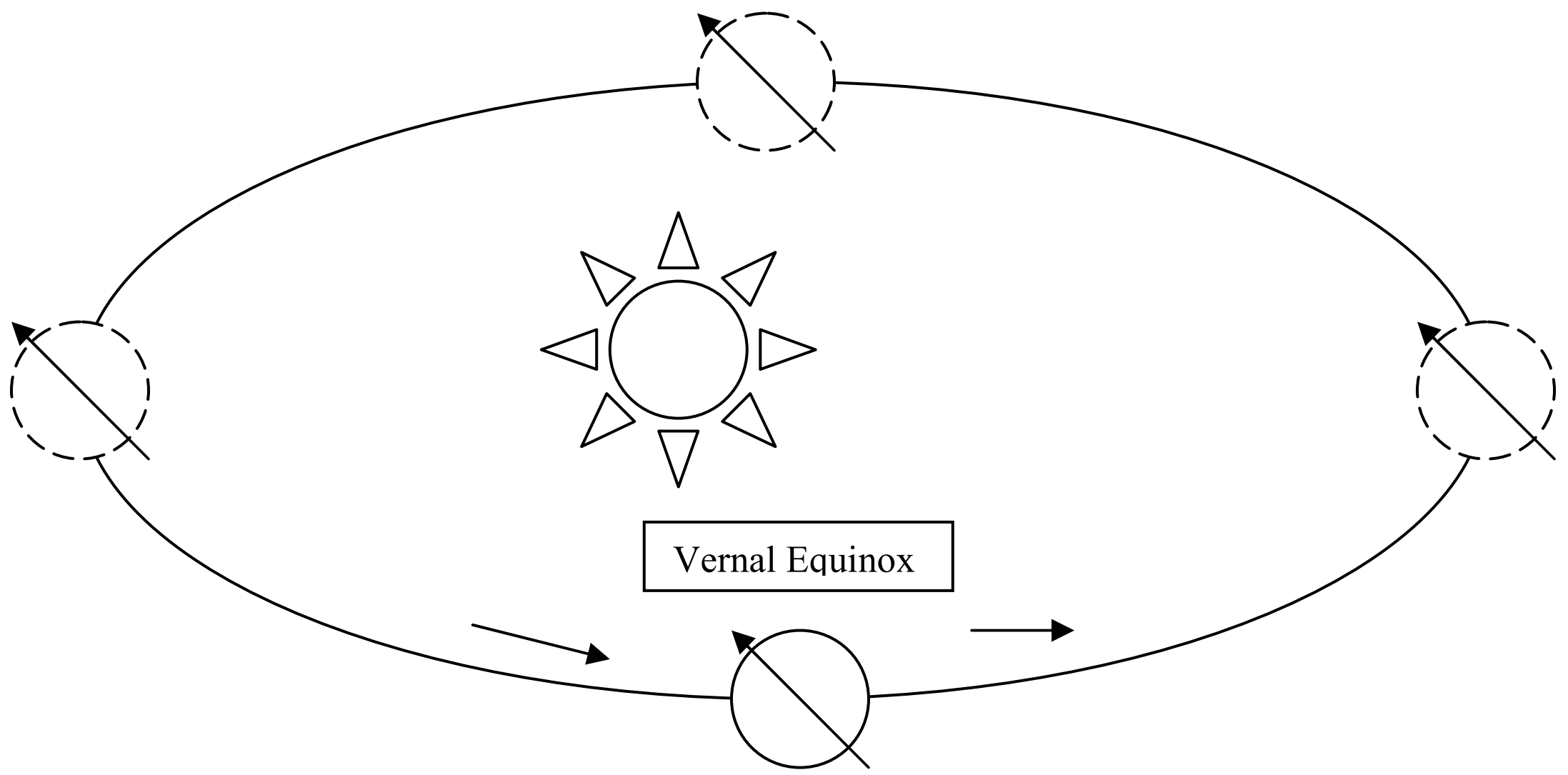

The name refers to the changes of seasons (greek ` τροπαι' or tropai, the turning points) which are fixed in this kind of year. It is for this reason that the tropical year is of great importance in the construction of calendars. Currently the tropical calendars are the most widely followed across the world.

#### 2.3.3 Siderial Year

A siderial year is the interval between two successive passages of the Mean Sun at the same (fixed) star. It lasts 365.256366 days.

#### 2.3.4 Anomalistic Year

An anomalistic year is the interval between two successive passages of the Earth through the perihelion (the point closest to the Sun) of its orbit and it lasts 365.259636 days.

### *2.4 Precession of Equinoxes*

This is the reason behind the difference in lengths of the Siderial and the Tropical year. In the Western world the Greek astronomer Hipparchus discovered it in 128 BCE.

Siderial Zodiac is defined by positions of the stars with origin at the $1^{st}$ point of Constellation Aries. On the other hand, the Tropical Zodiac is defined by the Vernal Equinox point, which is the intersection of Celestial Equator and the Ecliptic. This point by definition is the Tropical $1^{st}$ point of Aries.

In absence of precession both points should be fixed in the sky. If they are different points to begin with, then the distance between them should be seen as being constant from year to year.

Hipparchus compared the position of these points across the centuries after studying the old and new observational data and discovered that this difference was not constant. The Vernal Equinox point was found to be moving against the background of fixed stars. In other words, under Precession, the zodiac moves from West to East over a period of time as seen from the Earth and the difference between the Siderial and Tropical 1st point of Aries keeps on increasing. What is called Aries today is different from what was called Aries in the past in the Tropical Zodiac.

Vernal Equinox point in 290 CE occurred between Siderial Aries and Pisces as shown below. The observer is facing north.

```
      Pisces          VE      Aries
|----------------------*-----------------------|     (Ecliptic)
```

After sometime in future the relative position of VE and zodiacs changes and VE occurs in Pisces as it has moved westward relative to the fixed zodiac. Currently accepted value of the rate of precession is 50.29'' per year.

```
      Pisces          VE      Aries
----|-------------------*----|------------------------|         (Ecliptic)
```

Similar past positions of the Sun at different times of the year are give below.

**Solar positions through history**

| Solar position | 6650 BCE – 4400 BCE | 4400 BCE-2250 BCE | 2250 BCE- 100 BCE | 100 BCE-2050 CE |
|---|---|---|---|---|
| Vernal Equinox | Gemini | Taurus | Aries | Pisces |
| Spring Solstice | Virgo | Leo | Cancer | Gemini |
| Autumn Equinox | Sagittarius | Scorpio | Leo | Virgo |
| Winter Solstice | Pisces | Aquarius | Capricorn | Sagittarius |

The physical reason for Precession was not understood correctly for almost 2000 years. It had to wait until the application of the law of gravitation (discovered by Newton) to the Earth's rigid-body motion. Sun exerts a torque on the equatorial bulge of the Earth because of its tilt with respect to the Ecliptic. As a result the tip of the Earth's rotational axis executes a circle against the backdrop of the stars in the sky.

The precession has a period of about 25,700 years. Equivalently, the movement is one degree about every 72 years. It is not perceptible in an average lifetime but becomes apparent over a few centuries. It is not known if the rate of precession is constant or it varies over long time-periods.

### *2.5 Nutation*

The gravitational attraction between Sun, Moon, and Earth is dynamic as it changes with their relative orientation. Also the lunar orbit makes an angle of 5 degrees with the ecliptic. One also has to add that Earth and Moon are not quite spherical. When these effects are taken into account, the resulting motion has many frequency components out of which the precession is the dominant one. The next important motion is called **Nutation**, which appears as a wobble of the Earth's rotational axis superimposed on the precession. It is a motion with amplitude of 17 arc-seconds and a period of 18.6 years.

This has almost negligible effect on the calendar and has been considered only in modern times.

### *2.6 Planetary Motions*

The solar system is composed of other planets, comets and asteroids which move around and affect each other's motion in very complex ways. It is well known that the solar system as a whole is a stochastic system and its motion can not be predicted with certainty over a very long period of time. The current Newtonian approach supplemented with some General Relativistic considerations is sufficient to predict the solar system's motion over short term though. So we can safely ignore the effect of other solar bodies for the sake of the calendrical considerations.

# 3. Types of Calendars

The astronomical periods related to the Sun and Moon form the basic set of data for calendars. A good sourcebook is the “Explanatory Supplement to *Astronomy Almanac*, 1992, by Ken Seidelmann.

The basic astronomical units of day, month and year are the building blocks for all calendars. It is easily seen that the year does not contain an integral multiple of months, nor the month contains an integral multiple of days. In fact all these periods are incommensurate with respect to one another. This gives rise to many possibilities for defining their relations and that forms the basis of various calendars. A particular Calendar uses the above values to arrive at a practical scheme to divide the sequence of days and nights into manageable units of year and months.

This chapter contains the basic definitions and examples (non-Indian) of the three types of calendars. The examples based on the Indian calendars are given in the following chapter.

### *3.1 Lunar Calendar: Structure*

The Moon is the most conspicuously visible object in the sky, so the earliest calendars in history were based on its motion.

The basic period of this calendar is the Lunar Month (LM) defined as the time interval between two successive full moons and it is 29.530589 days. Twelve such LMs make a Lunar Year (LY) of 354.530589 days. This falls short of the Tropical Year by about 11 days, so very soon a mismatch develops between the LY and the seasons.

The two kinds of such calendars are natural and compensated. Natural lunar years are the simplest ones, which probably were the earliest calendars in civilization. They just follow the lunar motion.

The most famous example of the second kind is the Islamic Calendar. This scheme has religious sanction in the Islamic world as it was chosen by the prophet Muhammad himself to be followed by the Islamic community. In this calendar the day begins with the sunset on the evening preceding the same day according to the secular calendar.

For religious purposes, the start of a new month is not determined by the tables of the fixed calendar but through actual observation of the young moon's crescent.

### *3.2 Lunar Calendar: Example*

The basic structure of the Lunar Year is determined by the distribution of 354 days into 12 months. For the Islamic Lunar Year (ILY) this is given through the following rules.

(i) The Islamic Lunar Months (ILM) are either 30 or 29 days long.

(ii) There are six ILMs of either kind to form one ILY of 354 days. This makes the average length of the ILM as 29.5 days.

(iii) As the mean ILM is shorter than the true LM, a compensation scheme is still needed to keep the calendar synchronized with lunar motion as seen in the sky. Islamic calendar adds 11 days in 360 months (1 d in 2500 years discrepancy) resulting in an average of 29.5 + (11/360) = 29.530556 d. This should be compared with the accepted modern value of 29.530589 d.

(iv) An Embolismic Year (year with intercalation) contains 355 days. A common year has 354 days. In a 30 year cycle the year numbers 2, 5, 7, 10, 13, 16, 18, 21, 24, 26, and 29 are Embolismic Years.

(v) But as one can verify easily, it will still necessitate extra corrections over a longer period of time because the ILY is shorter than the true LY by 0.000033 days.

The month names are: muharram, safar, rab'ia I, rab'ia II, jumada I, jumada II, rajab, sha'ban, ramadan, shawwal, dhu al-q'adah and dhu al-hijjah)

For secular purposes a tabulated lunar calendar is used with an ordinary year with 354 days and 12 months that have alternating lengths of 29 and 30 days. In a cycle of 30 years, 11 leap years with 355 days appear in which the twelfth month has 30 instead of 29 days. There are, however, two different structures of the 30-year cycle in use, which cause differences in the date by one day during 348 of the 360 months. In either case the new year's day of the fixed Islamic calendar moves through all seasons within 33 years.

### *3.3 Solar Calendar: Structure*

The solar calendar is the most prevalent in modern times. Its origins in the western world go back to the Greek and Roman eras. The basic idea is to take the Sun as the primary object in determining the calendar. In ancient world, it took some time for observational astronomy to develop so the concept of solar year was born much later. In this scheme, a year is taken as the interval between two successive equinoxes or solstices and its modern accepted value is 365.2421875 days. The basic problem of the solar calendar revolves around fitting this number in a practical cycle of months. An important astronomical phenomena known as the precession of equinoxes complicates the matters further.

The two kinds of Solar Year are known as Siderial and Tropical.

(i) Siderial Year is the period after which the Earth's axis returns precisely to the same position in the FSF. In practical terms, the Siderial Year starts when the Sun is in at a particular position in the fixed frame of stars, e.g. 1$^{st}$ point of constellation Aries.

(ii) Tropical Year is the time between successive passages of the Sun through the 1$^{st}$ point of the Aries as seen on Earth (equals the time between start of seasons in consecutive years). Usually, the year starts when the Earth is at a specific point in its orbit with respect to the Sun, e.g. Spring Equinox in Iranian Calendar.

The time taken by the Sun to move from one cardinal point of the year to the other is as follows. (i) Spring Equinox to Summer Solstice: 92.8d. (ii) Summer Solstice to Autumn Equinox: 93.6 d. (iii) Autumn Equinox to Winter Solstice: 89.8 d. (iv) Winter Solstice to Spring Equinox: 89.0 d.

---

## *3.4 Solar Calendar: Examples*

### 3.4.1 Julian Calendar

The Julian calendar is based on a solar year with 365 days. To account for the fact that the tropical year is longer than 365 days by about a quarter day, a leap day is inserted at the end of month of February in every fourth year. This simple leap year rule was already known in late Egypt. It was in fact an Egyptian scholar from Alexandria named Sosigenes who advised Julius Cesar during the introduction of the calendar into the Roman empire in the year 46 BC. The calendar is named after Julius Cesar.

Julius Cesar had to start the introduction of his calendar with an anomalous leap year with 445 days for the year 46 BC to compensate for the inaccuracies of the Roman calendar used before. The following year 45 BCE was a normal leap year with 366 days. After Cesar's death the new leap year rules were at first incorrectly applied and too many leap years occurred. This was corrected under the government of Augustus and the Julian calendar was strictly obeyed since the year CE 8. For earlier years date estimates are uncertain by a few days since the sequence of leap years is not exactly known.

In astronomy and for historical purposes the Julian calendar is also applied to epochs earlier than the year 46 BCE when this calendar was not yet defined and the people of that time could not know their date in it. To indicate this extension, the term *proleptic Julian calendar* is occasionally used (proleptic = brought forward).

---

### 3.4.2 Gregorian Calendar

The Julian year with its duration of 365.25 days is longer than the real tropical year by 0.0078 days or 11 minutes 14 seconds. Although imperceptible over short duration, it accumulated over the centuries, and was noticed when the true Vernal Equinox was found to have moved away from the March 21, the beginning of Spring as decreed by the Catholic church. The difference was about 10 days at the beginning of the $16^{th}$ century.

To solve this problem, pope Gregory XIII in CE 1582 ordered a calendar reform for the domain of the Catholic Church. Its main pints were the following.

1. *Omission of 10 calendar days*: The 4th of October 1582 was followed directly by the 15th of October 1582 in the new calendar. This

brought the start of spring back to the 21st of March. The reckoning of weekdays was not changed.

2. *Introduction of a new leap year rule*: No leap days will occur in years divisible by 100 but not by 400. This results in a mean year length of 365.2425 days. The remaining difference with respect to the tropical year is small enough to require the insertion of an extra leap day only after 3333 years. The leap day is inserted at the end of February as in the Julian calendar.

The objectives and details of the new calendar were described in AD 1603 by Christoph Clavius in his book ``Explication Romani Calendarii a Gregorio XIII P.M. restitui".

The non-Catholoc countries hesitated to adopt the new calendar, in some cases for very long times. Turkey, for instance, converted to the Gregorian calendar on January 1, 1927.

The start of the year count of the Common Era was fixed in Jesus's supposed birth year by Dionysius Exiguus in 525 CE. Apparently he made a mistake because currently Jesus is supposed to have been born in 4 BCE.
In this counting the year AD 1 is directly preceded by the year 1 BC, a year 0 does not exist in this system.

In contrast, the astronomical reckoning indeed uses a year 0. For the purpose of distinction, astronomical reckoning drops the symbols AD and BC and uses a plus or minus sign before the year instead. The astronomical year +1 therefore corresponds to the year CE 1, the year 0 corresponds to 1 BCE, and the year -1 to 2 BCE.

The Gregorian calendar is regularly used in astronomy for dates later than October 14, 1582. For some applications, however, it is favorable to extrapolate it to epochs before this date (so-called proleptic Gregorian calendar). On the other hand, even today some dates or time intervals are calculated according to the Julian calendar.

### 3.4.3 Pre-Islamic Egyptian Calendar

Since the fourth millennium before Christ a solar year with a length of 365 days was used. The year was divided into 12 months with 30 days each, plus five additional days. The months were combined into groups of four months each to form the flooding, seeding, and harvesting seasons, referring to the yearly floods of the river Nile. The relation of these seasons to the beginning of the Egyptian calendar year was variable, though, because on average the Nile flood appears at the same time of the tropical year and consequently seeding and harvesting have to follow in step. The start of the seasons was therefore defined by the heliacal rising of the star Sirius (the Egyptian name of which was Sothis).

The heliacal rising is the first rise of a star visible in the pre-dawn after its conjunction with the sun. Strictly speaking, the heliacal rising does not define the length of a tropical, but of a siderial year if the star's proper motion can be ignored. However, the difference was insignificant for Egyptian time keeping.

The Egyptian calendar made no use of leap days, so in a period of 1460 years the new year's day moved through all seasons. For the Egyptians, however, it appeared as if the heliacal rising of the Sothis (or Sirius) moves with this period through the calendar. It was therefore called the Sothis cycle.

### 3.4.4 Iranian Calendar

Pre-Islamic Iran followed the religion of Zoroastrianism and had a solar calendar. The calendar survived the onslaught of Islam and has become a part of the Iranian national identity. Its present form was established in 11$^{th}$ century by a panel of astronomers appointed by Jalal-el-din Seljuq and so it is also known as Jalali calendar. The panel included the famous poet and astronomer Omar Khayyam.

The calendar is based on the tropical solar year and starts on the midnight closest to the Vernal Equinox. The Persian month names have been retained.

**Month Names of Iranian Calendar**

| Month | Begins on | Month | Begins on |
|---|---|---|---|
| Farvardin (31) | Vernal Equinox | Mehr (30) | Autumn Equinox |
| Ordibehesht (31) | | Aban (30) | |
| Khordad (31) | | Azar (30) | |
| Tir (31) | Summer Solstice | Dey (30) | Winter Solstice |
| Mordad (31) | | Bahman (30) | |
| Shahrivar (31) | | Esfand (29/30) | |

A complex method for inserting leap years in a cycle of 128 years was devised. This results in one missing day from the solar cycle in 141,000 years. The resulting accuracy is better than that of the Gregorian or Common Era calendar, which misses one day in 5,025 years.

## *3.5 Lunisolar calendars: Structure*

A luni-solar year attempts to have a solar calendar but instead of a solar day it uses the lunar month as the building block. It tries to combine the phases of the moon and the seasons into one calendar. This is possible if leap months are inserted. Several schemes were used in history.

The basic period of this calendar is the Lunar Month (LM) defined as the time interval between two successive full moons and it is 29.530589 days. Twelve such LMs make a Lunar Year (LY) of 354.530589 days. As we can see, this falls short of a Tropical Year of 365.242190 days (=12.368266 LM) by about 11 days. If not compensated, the LY will not keep a constant relation with the seasons. Many schemes of keeping the LY and TY in synchronization were developed later and they form the basis of lunisolar calendars.

### 3.5.1 Method of Cycles

The most famous such scheme is better known as the Metonic cycle after the Greek astronomers Meton and Euctemon (432 BCE) but apparently it was also known to other cultures before.

The Metonic cycle encompasses a total of 235 months of which 125 are ‘full’ (i.e. they have 30 days) and 110 are `hollow' (having 29 days). The months are combined into 12 normal years with 12 months each and 7 leap years with 13 months each. The cycle covers 6940 days whereas 235 synodic months sum up to 6939.688 days and 19 tropical years to 6939.602 days. The difference in motion between Sun and Moon amounts to only 0.0866 days so that eclipses repeat in the Metonic cycle with high accuracy.

The intercalation scheme follows the following rules.

(i) It has cycle of 19 lunar years with 12 months each, and 7 extra months in year number 3, 5, 8, 11, 13, 16, and 19. So the total number of months is 235 (= 228 + 7), in which 125 have 30 and 110 have 29 days (=6,940 days). Regular 235 months (each having 30 days) have 7050 days, so 110 days need to be dropped.

(ii) One day is dropped every 64 days and the month in which it occurs is made deficient (29 d).

(iii) Average duration of the month becomes 29.5319 d (true value = 29.530589 d).

(iv) Average duration of the year is 365.2632 d (true value = 65.242190 d).

### 3.5.2 Improvements over Metonic Cycle

Many improvements were proposed by the later astronomers to improve upon the Metonic cycles.

(i) Calippic cycle (Calippus of Cyzicus, 370-300 BCE):

It has 76 years ( or 4 Metonic cycles) with 1 day omitted every $4^{th}$ Metonic cycle. Average duration of the month = 29.5308 d (compare with true value = 29.530589 d),

Average duration of the year = 365.242190 d).

(ii) Hipparchic cycle (Hipparchus of Nicea, 180-125 BCE):

It has 304 years (=4 Calippic cycles) with 1 day omitted every $4^{th}$ Calippic cycle. Average duration of the month = about 0.5 sec less than true Synodic cycle. Average duration of the year = about 6 minutes longer than true year.

### 3.5.3 A Mathematical Model for Intercalation

The basic mathematics of intercalation can be understood by the following.

Let

R = average number of months to be intercalated per year,

Y = tropical year duration, and

L = Lunation.

Then

R = (Y/L) – 12 = 0.368270 (modern value)

Successive rational approximations to this are 1/3, 3/8, 4/11, 7/19, and 123/334. For 100 and less years 7/19 (Metonic cycle) can not be improved. The Jewish, Hindu, Chinese, Zoroastrian, Buddhist, Jain, and Sikh religious calendars all of them have incorporated the Metonic cycles in one way or another. It should be noted that the Indian way of intercalation is unique and is an original contribution in this area.

### 3.5.3 Intercalation in India

The Indian astronomers based their calculation on *tithi*s.

1 lunar month = 29.5 d (more accurately 29.530588 d) = 30 *tithi*s

1 solar month = (365.25/12) d = 30.4375 d

= 1 lunar month + 0.9375 d

In every 32.5 (more accurately 32.46) solar month, we have 33.5 lunar month. This excess is called *adhika māsa* or intercalary month.

Another equivalent method is indicated in *vedāṅga jyotiṣa*.

1 day = (360/354) *tithi*s = 1.0169 *tithi*s

1 solar year = 365.25 X 1.0169 *tithi*s = 371.09 *tithi*s

This extra number of *tithi*s is called *Ṛtu-śeṣa*.

The number of *tithi*s in 19 solar years = 19 X 371.09 = 7049.95

The number of *tithi*s in 19 lunar years = 19 X 360 = 6840

The difference = 210 = 7 lunar months.

This is nothing but Metonic cycle in language of *tithi*s. An ancient scholar *lagadha* has mentioned a 19 year *yuga* cycle with regular 18 years of 371 *tithi*s and the $19^{th}$ year of 372 *tithi*s. Even longer cycles of 95 years were observed.

## *3.6 Lunisolar Calendars: Examples*

### 3.6.1 Chinese Lunisolar Year

Traditional Chinese calendar is a luni-solar one. It uses lunar months to describe a tropical solar year, which starts from the Winter Solstice in December.

In this calendar, the day starts at midnight and the month starts on the New Moon's day. The calculations are based on the longitude of the 120 degrees east. Also the Chinese New Year falls on the second new moon after the December in general. There are some exceptions to this general rule. The necessary intercalation of leap months lead to the development of the Metonic cycle with 7/19 rule.

In early times the years were not counted but 60 of them were grouped together. This sexagenary cycle uses 10 heavenly stems (tian gan) and 12 earthly branches (di zhi). The heavenly stems are: Wood, Fire, Earth, Metal and Water. Each is repeated two times and denoted by two different characters in Chinese orthography. The earthly branches are named after the animals. They are: Rat, Ox, Tiger, Rabbit, Dragon, Snake, Horse, Sheep, Monkey, Rooster, Dog, and Boar.

A year has a unique designation based on these two identifiers. If we denote them by numbers then the sequence is given as (1,1), (2,2),…, (10,10), (1,11), (2,12), (3,1), (4,2),.., and so on until 6 cycles of stems and 5 cycles of branches are all used. This gives a cycle of 60 years. In ancient China, these 60-year periods were named according to an important event or a sovereign of that epoch.

The Chinese claim that the counting of this era started with the mythical emperor Shih Hunag Ti. In that case this may be the longest continuously used era in history.

### 3.6.2 Pre-Christian Greek Lunisolar Year

In old Greece a lunisolar year was used, with intercalation rules that were in the beginning primitive and irregular. Since about 500 BC the *octaeteris* gained wide-spread acceptance, a rule with 8-year cycle in which five ordinary years with 12 months each are combined with three leap years of 13 months each. In the year 432 BC, Meton in Athens found the 19-year cycle named after him (although it was discovered independently in other cultures). Of similar quality, although longer in period and therefore more difficult to use, was the Callipic cycle that equated 76 years with 940 months and 27759 days.

### 3.6.3 Jewish Lunisolar Year

The year reckoning of the modern Jewish calendar begins with the year 3761 BC when according to the Jewish creed the world was formed. This reckoning was established in about the 10th century AD, the calendar itself found its final form already in the 4th century AD. The calendar is based on a lunisolar year with a complicated set of intercalation rules. The particular complexity is the consequence of an attempt to avoid certain feasts to fall on week days considered improper. Therefore, one distinguishes `defective', `normal', and `complete' ordinary years with 353, 354, and 355 days, respectively, and corresponding leap years with 383, 384, and 385 days.

The day commences at 18 o'clock in the Jewish calendar. This is a common characteristic of lunar calendars since the moon's slim crescent after the new moon is visible shortly after sunset. With it begins a new month and thus also a new day.

## *3.7 Non-Astronomical Calendars*

There are many non-astronomical calendars followed by various ethnic and religious groups. A few of them are described below.

- *Latin America (historical):* The advanced cultures in Latin America used a ritual calendar with a period of 13 times 20 days in combination with a solar year that consisted of 18 months with 20 days each plus five extra days (which were considered calamitous). This resulted in a 52-year cycle. In general, there was no continuous count of the years. Only the Maya counted the years, starting from September 6, 4113 BC in units of 'kin' (1 day), 'vinal' (20 days), 'tun' (360 days = 18 vinals), 'katun' (7200 days = 20 tuns) and 'baktun' (144000 days = 20 katuns).

- *Calendar of the French revolution*: This calendar was designed by S. Marechal in 1787 and established in post-revolutionary France on October 5, 1793. Its first year began (nominally) on September 22, 1792, and new years started on the astronomically determined autumn equinox. The year was divided into 12 months with 30 days each, to which were added five or six extra days (the 'sansculotides'). Each month consisted of three decades of 10 days length, the day was divided into 10 hours, the hour into 10 parts and so on.

  This calendar did not catch on and so the Gregorian calendar was re-established in France on January 1, 1806.

- *Civilian calendar of the Federal Republic of Germany*: With typical German thoroughness, this calendar is standardized in the norm DIN 1355. It defines the length(s) of the year, the leap year rules, the names of months and week days, the suffixes *vor Christus* and *nach Christus* (for BC and AD, respectively), and the reckoning of years and weeks. These specifications are in general agreement with the Gregorian calendar and add only items unspecified by the Gregorian calendar.

- *The Baha'I Calendar*: The year (also called Badi) is a tropical solar year and it consists of 365 days (366 in leap year). The days are organized in 19 months of 19 days each. The extra 4 days (5 in leap years) fall between the 18th and 19th months. These "intercalary days" are observed as days of rest (Ayyám-i-Há).

Each month is named after an attribute of God in Islamic tradition:

Bahá (Splendour), Jalál (Glory), Jamál (Beauty), 'Azamat (Grandeur), Núr (Light), Rahmat (Mercy), Kalimát (Words), Kamál (Perfection), Asmá' (Names), 'Izzat (Might), Mashíyyat (Will), 'Ilm (Knowledge), Qudrat (Power), Qawl (Speech), Masá'il (Questions), Sharaf (Honour), Sultán (Sovereignty), Mulk (Dominion), and 'Alá (Loftiness) (month of fasting similar to Ramadhan).

The days are also named after those of God: Jalál (Glory) = Saturday, Jamál (Beauty) = Sunday, Kamál (Perfection) = Monday, Fidál (Grace) = Tuesday, 'Idál (Justice) = Wednesday, Istijlál (Majesty) = Thursday, and Istiqlál (Independence) = Friday (day of rest).

As is the case with Jewish and Islamic reckoning, the day begins at sunset. The March 21, 2005 CE is the start of 161 BE (Baha'i Era).

- *The Pawukon calendar of Bali*: The island of Bali in Indonesia follows three calendars, the usual Gregorian, Indian lunisolar Shaka and Pawukon. The last consists of 210 days and is divided not according to weeks and months but according to ten separate week systems running concurrently. These contain 1, 2, 3,…, 10 days and are known by their Sanskrit-based names.

**Day Names of Bali Calendar**

| **Week Names** | **Day Names** |
|---|---|
| Ekawara | Luang |
| Dwivara | Menga, Pepet |
| Triwara | Pasah, Beteng, Kajeng |
| Caturwara | Sri, Laba, Jaya, Menala |
| Pancawara | Umanis, Paing, Pon, Wagé, Kliwon |
| Sadwara | Tungleh, Ariang, Urukang, Paniron, Was, Maulu |
| Saptawara | Redite, Coma, Anggara, Buda, Wraspati, Sukra, Saniscara |
| Astawara | Sri, Indra, Guru, Yama, Ludra, Brahma, Kala, Uma |
| Sangawara | Dangu, Jangur, Gigis, Nohan, Ogan, Erangan, Urungan, Tulus, Dadi |
| Dasawara | Pandita, Pati, Suka, Duka, Sri, Manuh, Manusa, Eraja, Dewa, Raksasa |

A given day has 10 names according to these 10 systems. Their combinations signify good and bad days for particular activities like construction, fishing, studies, etc and also for some religious activities. This calendar does not have an associated era or year count. The "Saptawara" system corresponds to the usual week and each of the 30 such weeks have their own names.

**Week Names of Bali Calendar**

| 1-5 | 6-10 | 11-15 | 16-20 | 21-25 | 26-30 |
|---|---|---|---|---|---|
| Sinta | Gumbreg | Dunggulan | Pahang | Matal | Ugu |
| Landep | Wariga | Kuningan | Krulut | Uye | Wayang |
| Ukir | Warigadian | Langkir | Merakih | Pujut | Kelawu |
| Kulantir | Julungwangi | Medangsia | Tambir | Prangbakat | Dukut |
| Taulu | Sungsang | Pujut | Medangkungan | Bala | Watugunung |

Balinese specify a day by giving its week and day names. As an example 'Redite Uye' is the Sunday of the 22nd week.

# 4. Indian Calendars

Indians have used many kinds of calendars during their long history. The beginnings of the astronomical ideas can be traced to the Vedas and the earliest treatise of *vedān˙ga jyotiṣa* is aware of the ingredients of the calendar. Out of all ancient cultures, India alone did not have one official calendar across the whole country due to historical reasons. This was somewhat mitigated by the fact that these differences amounted to variations on a few basic templates.

Here the phrase "Indian calendar" will be used for the pre-Islamic calendars. A few new calendars came into existence during the Islamic rule but they did not change the earlier approaches in any substantive way.

## *4.1 Traditional (Siderial Solar)*

Indian solar calendars are based on Siderial Year (SY), in which the Precession of Equinoxes is not taken into account. The year begins when the sun reaches a certain fixed point in the ecliptic. At present, this point is marked by a star in Pisces, which lies about $28^0$ before the current vernal equinox. At the beginning of the calendar the vernal equinox coincided with this point.

The traditional solar year was divided in twelve months (*māsa*) and also in six seasons (*Ṛtu*), each season consisting of two months. The names of months were those given by the zodiac occupied by Sun during that month. Their names and duration (in days) are given in the table below.

**Month Names of the Indian Solar Year**

| **Solar Month Names** | **Actual Duration (days)** | **Common year equivalent** | **Comments** |
|---|---|---|---|
| *Meṣa* | 30.9 | April-May | |
| *Vṛṣa* | 31.4 | May-June | |
| *Mithuna* | 31.6 | June-July | |
| *Karka* | 31.5 | July-August | *karka-saṃkrānti* occurs on July 14. Start of *dakṣiṇāyana* half of the year. |
| *Siṃha* | 31.0 | August-September | |
| *Kanyā* | 30.5 | September-October | |
| *Tulā* | 29.9 | October-November | |
| *Vṛśchika* | 29.5 | November-December | |
| *Dhanu* | 29.4 | December-January | |
| *Makara* | 29.5 | January-February | *makara-saṃkrānti* occurs on 14th January. Start of the *uttarāyaṇa* half of the year. |
| *Kumbha* | 29.8 | February-March | |
| *Mīna* | 30.3 | March-April | |

The months are the number of days taken for the Sun to travel from one zodiac to another in the fixed stars frame. These months have unequal values due to the ellipticity of the earth's orbit. The number of days in each month varies from 29 to 32 days. The *mithuna* is the longest and the *dhanu* is the shortest month as seen in the table above.

Actually, a solar month is named after the *amānta* lunar month, which begins in it, which, in turn is named after the *nakṣatra* which coincides with the full moon in it, and so the *nakṣatra* that `marks' the vernal equinox is after it by up to about 15°.

The siderial year in this system continually advanced on the tropical year because it is longer than Tropical Year (by 20 minutes and 25 seconds) due to precession of the equinoxes. This amounts to about (i) one day in 71 years, or (ii) about 1 *nakṣatra* every 960 years, or (iii) roughly one month every two millennia.

The correspondence of this calendar with the Gregorian one varies from year to year; though the overall drift is small (about 1 day every 71 years). The more important difference between the two is the following.

---

- Both calendars count the years astronomically.
- In the Gregorian calendar, the lengths of the months are based on some rules and so they are conventional. This means that the beginning or end of the months like January, are not marked by any astronomical phenomena.

---

- In the sidereal calendar the length of the months are determined according to the observed motion of Sun across the zodiacs. Currently though Indian sidereal calendar calculates them according to a model of the solar system given in *sūrya siddhānta*.

---

Only the tropical year keeps correct synchronization with the seasons, so the small difference of the sidereal year from the tropical one gets magnified over the centuries. Right now it is out of synch with the seasons and there is a discrepancy of about three weeks between the start of seasons indicated by *saṃkrānti* months and the actual seasons themselves.

The most widely noticed such discrepancy is experienced around January 14th of every year. According to the *panchān˙ga* (the Indian almanac), the Sun becomes *uttarāyaṇa* on that day, which means that it has reached the southernmost point in its orbits as seen from the Earth in the east and has started moving northwards. But astronomically that takes place on the Winter Solstice day of usually Dec 22 of the previous year. This suggests that this solar year was devised somewhere around 285 CE. One way out of this discrepancy is to say that the word *uttarāyaṇa* denotes the entry of the Sun in Siderial Capricorn only and it does not signify the northward solar motion.

In ancient times, attempts have been made to start the year (and the counting of the *nakṣatras*) near one of the equinoxes. This leads to the `first' *nakṣatra* marking the vernal equinox shifting back from *mṛgaśirā* to *rohiṇī* to *kṛttikā*, and ultimately to the current usual counting starting with *aśvinī*. As the apparent speed of the sun varies during the year (currently faster during the current winter in the northern hemisphere), and also slightly from year to year, the number of days in a month varies between 29 and 32 days. Usually now, the northern hemisphere winter months are 30 days and the summer months are 31 days each.

Incidentally, a solar day in Indian Calendar is counted strictly from sunrise to sunrise.

The sidereal solar year has regional variations, which occurred due to different historical developments and a lack of central religious and political authority.

(i) Traditional Solar-1 (Tamil Nadu)

The year starts on the *meṣa saṃkrānti* (Sun-crossing into siderial Aries) in Mid-April. The month names are: *chittirai, vaikāsi, āni, āḍi, āvaṇi, puraṭṭāsi, aippasi, kārtikai, mārgazhi, tai, māsi*, and *panguni*. These do not follow the traditional Sanskrit names except a few. It should be noted that the year starts in *chittirai* (a Tamil cognate of lunar month *chaitra*), which actually signifies a solar month.

The beginning of a solar month is determined by the following rule.

- If the *saṃkrānti* occurs before the sunset, then the month begins on that day.
- If *samkrānti* occurs after the sunset, then the month begins on the next day.

The years have individual names in a sixty years cycle named after the Jupiter. It takes Jupiter about twelve years to orbit the Sun, so the cycle contains five Jovian years. The Chinese calendar also follows a similar cycle of sixty years. Given the traditionally close relations, there may have been borrowing of ideas between India and China for this to happen.

The names of the weekdays are: *nāvirruk-kizhamai* (Sunday), *tingạt-kizhamai* (Monday), *chevvāy-kiZhamai* (Tuesday), *putan-kizhamai* (Wednesday), *viyāzhak-kizhamai* (Thursday), *veḷḷik-kizhamai* (Friday), and *chanik-kizhamai* (Saturday). Many of these names are from ancient form of the Tamil language.

(ii) Traditional Solar-2 (Orissa)

The year starts on the *meṣa saṃkrānti* (Sun-crossing into Siderial Aries) in Mid-April.

The beginning of a solar month is the same day when *saṃkrānti* occurs. This rule is also followed in Panjab and Hariyana.

(iii) Traditional Solar-3 (Assam, West Bengal, and Tripura)

The year starts on the day after the *meṣa saṃkrānti* (Sun-crossing into Siderial Aries) in Mid-April. Some details about the Bengali version are given below.

The month names are: *vaiśākha, jyeṣtha, āṣāḍha, śrāvaṇa, bhādrapada, āśvina, kārttika, agrahāyaṇa, pauṣa, māgha, phālguna*, and *chaitra*. These names apply to solar months and should not be confused with similar names in the lunar calendar. The months were originally named by a neighboring *nakṣatra* when the moon was full. The first month of *vaiśākha* starts on that day on whose sunrise the apparent sun's longitude is first past 23 degrees and 15 minutes (thus entering *meṣa*). The months are always a whole number of days, but they vary from year to year in length.

The beginning and end of a month is decided by the apparent position of the sun within this zodiac (each month lasting 30 degrees). This is calculated by using formulas based on the fifth century astronomical treatise *sūrya siddhānta*, updated by the great Bengali astronomer *Rāghavānanda Chakravartī* . His method is given in two treatises called *siddhānta rahasya* and *dinachandrikā* dating to 1591 and 1599 CE respectively.

The beginning of a solar month is determined by the following rule.

- When the *saṃkrānti* occurs between the sunrise and the following midnight, then the month begins on the next day.
- If *saṃkrānti* occurs after the midnight, then the month begins on the day following the next day, i.e. on the third day.

(iv) Traditional Solar-4 (Kerala).

The Kerala year starts on the *siṃha saṃkrānti* (Sun crossing into sidereal Leo) in August-September. The month names are: Chingom (*siṃha*), Kanni (*kanyā*), Thulam (*tulā*), Vrischikom (*vṛśchika*), Dhanu (*dhanu*), Makarom (*makara*), Kumbham (*kumbha*), Meenom (*mīna*), Medom (*meṣa*), Idavom (*vṛṣabha*), Mithunom (*mithuna*), and Karkidakom (*karkạta*). These are the Malayalam variants of the traditional Sanskrit *rāśi* (zodiac) names. The number of days in each month may vary from 29 to 32 days. For instance, the month of Makarom has 29 days while that of Karkidakom has 32 days.

The beginning of a solar month is determined by the following rule.

- If the *saṃkrānti* occurs before the *aparāhṇa* (before the $3/5^{th}$ of the time period between the sunrise and the sunset), then the month begins on that day.
- If *saṃkrānti* occurs after the *aparāhna*, then the month begins on the next day.

This calendar follows the Kollam era, which started with the establishment of Kollam city, in 825 A.D.

## *4.2 National Reformed (Tropical Solar)*

A reformed Indian calendar was established on March 22, 1957. Its year length and its leap year rules are the same as those of the Gregorian calendar, but the New Year's Day and the year count differ. The New Year's Day is the Vernal Equinox day that falls around March 21-22. For instance, March 22, 1957 corresponded to the beginning of the year 1879 in the historical *śaka* year reckoning, and in leap years the New Year's Day falls on March 21 of the Gregorian calendar. The table below gives the month names and their beginning dates

Months of the National Indian Calendar

| **Month** | **Begins on** | **Month** | **Begins on** |
|---|---|---|---|
| *Chaitra* (30/31) | 22 March | *Vaiśākha* (31) | 21 April |
| *Jyeṣtha* (31) | 22 May | *Āṣāḍha* (31) | 22 June |
| *Śrāvaṇa* (31) | 23 July | *Bhādra* (31) | 23 August |
| *Āśvina* (30) | 23 September | *Kārttika* (30) | 23 October |
| *Agrahāyaṇa* (30) | 22 November | *Pauṣa* (30) | 22 December |
| *Māgha* (30) | 21 January | *Phālguna* (30) | 20 February |

The month names are those of the traditional lunar ones but they mean tropical solar months in this calendar. This calendar did not become popular because its use of tropical year is very much against the Indian tradition of sidereal calendar.

## *4.3 The Nānakshāhī Calendar (Tropical Solar)*

The religion of Sikhism was founded by Guru Nanak in India in the 15th century. The followers are called Sikhs (meaning disciples) and are concentrated mostly in India. They in general followed the traditional Indian lunisolar calendar for fixing their religious festivals. Recently a group of Sikh scholars have proposed an alternate calendar based on the tropical solar year. It is tied to the Common Era Gregorian calendar in its structure. The apex Sikh religious body SGPC has accepted it, but still there are controversies concerning its adoption. Only the time will tell, if it will be accepted by the Sikh common man.

The correspondence between the *Nānakśāhī* and the Common Era calendar is given in the following table.

**Months of Nanakshahi Calendar**

| Month | Begins on | Month | Begins on |
|---|---|---|---|
| Chet (31) | 14 March | Vaisakh (31) | 14 April |
| Jeth (31) | 15 May | Harh (31) | 15 June |
| Sawan (31) | 16 July | Bhadon (30) | 16 August |
| Asu (30) | 15 September | Katik (30) | 15 October |
| Maghar (30) | 14 November | Poh (30) | 14 December |
| Magh (30) | 13 January | Phagun (30/31) | 12 February |

The month names are the Panjabi form of usual Indian names. In this calendar they denote solar months whereas traditionally these names stand for lunar ones. The beginning dates correspond more closely with the entry of the Sun in Siderial Zodiac, which is the same as in the traditional Indian Solar year. The length of the months has been fixed according to the tropical solar year.

### *4.5 Traditional Lunisolar Year*

We will use the term Indian Lunisolar year to denote an umbrella term to cover all the pre-Islamic and pre-Christian such calendars of South Asia. This system includes a complex interplay of lunar and solar cycles.

The religious festivals and holy days of Hindu, Buddhist, Jain, and Sikh religions are observed according to the lunar days.

The *saṃkrānti* or crossing of the Sun occurs at the time when the Sun is observed to enter one of the zodiacs. India follows the *nirayaṇa* or Siderial system for this purpose. In this system, no account is taken of the precession of the equinoxes. The fixed initial point on the ecliptic is the point opposite to the star *chitrā* (Spica or Alpha Virginis). The Siderial Aries starts from this point and each 30 degrees of the ecliptic afterwards corresponds to one of the zodiacs.

This fixed initial point was the Vernal Equinox day (*viṣuva*) in 285 CE. The current equinox point on the ecliptic has shifted by $23^{0}49'$ (on Jan 1, 1997) from this initial point due to the precession of the equinoxes. The starting points of other zodiacs have also shifted by the same amount relative to the equinox. The Indian calendar still retains the memory of its beginning so January 14th (*makara saṃkrānti*) is still referred to as *uttarāyaṇa*. In 285 CE, this coincided with the correct northern movement of the Sun, but currently this takes place around December 21.

The months are determined by the phases of the moon and there are two such systems.

(i) The *amānta* months end with the New Moon (*amāvasyā*).

(ii) The *purṇimānta* months end with the Full Moon (*purṇimā*).

---

Regional variations of this general approach give the following calendars:

---

(i) Lunisolar *amānta-1* (Andhra Pradesh, Goa, Karnataka, Maharashtra)

The lunar year (*chāndra varṣa*) begins on the day after the last New Moon day (*amāvasyā*) before the *meṣa saṃkrānti*. The first month is *chaitra* and it starts with the bright half (*śukla pakṣa*).

(ii) Lunisolar *amānta-2* (Gujarat)

The lunar year begins on the day after DeepAvalee or Diwali. The first month is *agrahāyaṇNa* and it starts with the bright half (*śukla pakṣa*).

In the Kathiawar area the year begins on *āṣā□ha śukla* 1.

---

(iii) Lunisolar *pūrṇimānta* (Bihar, Chhattisgarh, Delhi, Himachal Pradesh, Jammu & Kashmir, Jharkhand, Madhya Pradesh, Panjab, Rajasthan, Uttaranchal, and Uttara Pradesh)

This information is also given in the table below.

**Regional Variations of months in India**

| **Region** | **Month System** | **Year Beginning** |
|---|---|---|
| Maharashtra, Karnataka, Andhra Pradesh | | Beginning of *chaitra* (lunisolar year) |
| Kutch and Saurashtra (in Gujrat) | *amānta* | Beginning of *āṣāḍha* (lunisolar year) |
| Rest of Gujrat | | Beginning of *kārttika* (lunisolar year) |
| Tripura, Assam, Bengal, Tamil Nadu, Kerala | *amānta* (for religious dates) | Solar *chaitra* (sidereal solar year) |
| Bihar, UP, MP, Rajasthan, Haryana, | *Pūrṇimānta* | Middle of *chaitra* (lunisolar year). Same |
| Kashmir | | as beginning of *chaitra* in *amānta* system |
| Orissa, Punjab | *Pūrṇimānta* (for religious dates) | Solar *chaitra* (sidereal solar year) |

The lunar year begins on the day after the last full Moon day (*pūrṇimā*) before the *meṣa saṃkrānti*. The lunar month starts with the dark half (*kṛṣṇa pakṣa*). It is named after the *amānta* month, which starts one fortnight later. So a *Pūrṇimānta* month starts a fortnight earlier than the *amānta* month of the same name and ends in the middle of the same.

This system still uses *amānta* (called *mukhya-māna*) months to fix the religious festivals. It is to be noted that the year starts in the middle of the *Pūrṇimānta chaitra* month or from *chaita śukla* 1. The *chaita kṛṣṇa pakṣa* belongs to the previous year.

The start and end of the lunar days (*tithi*) are defined by requiring that the distance between the sun and the moon change by 12 degrees; so their actual length is between 20 and 27 hours. The *tithis* in a given month are named from *śukla* 1 through *śukla* 15 and then *kṛṣṇa* 1 through *kṛṣṇa* 15. The 14 days of each half (*pakṣa*) other than the new moon (*amāvasyā*) that ends the *kṛṣṇa pakṣa* or full moon (*purṇimā*) that ends the *śukla pakṣa* are usually referred to by the Sanskrit feminine ordinals: *pratipadā* (this word is an exception and does not mean first; it means beginning or entrance), *dvitīyā*, *tritīyā*, *chaturthī*, *panchamī*, *Ṣaṣthī*, *saptamī*, *aṣtamī*, *navamī*, *daśamī*, *ekādaśī*, *dvādaśī*, *trayodaśī*, and *chaturdaśī* (14th).

When solar days are named by a *tithi*, it refers to the *tithi* at the sunrise. Religious ceremonies also take into account the *nākṣatra* month (siderial lunar month determined by position of the moon in the zodiac).

## Intercalation

**(i) Extra Months**

As previously mentioned, the lunisolar calendars must add extra months according to some rules. Usually the Metonic cycle rules are followed by western calendars like Jewish. The Indian astronomers came up with a very ingenious method, which is equivalent to the Metonic cycle in the long run. It uses the true positions of the Sun and Moon. An intercalary month (*adhika-māsa*) is introduced whenever two lunar months start in the same solar month. In other words there are two new moons (in *amānta* system) between two consecutive *saṃkrānti* ocurrences. The name of the first of them is prefixed with the word *adhika*, and is intercalary. The second month is the normal one. The intercalary months occur at an interval of 2 years 11 months, or 2 years 10 months, or 2 years 4 months. The average interval is 2.7 years which is the same as given by the Metonic cycle.

(ii) Missing Months

Sometimes three solar months occur within one lunar month. This happens in winter during the solar months of *agrahāyaṇa*, *pauṣa*, and *māgha*. In this case no *amāvasyā* (new moon) will occur and so there will be no lunar month named after this solar month. This missing month or *kṣayamāsa* is always accompanied by two *adhika-māsa* months. One comes before and other after the *kṣaya-māsa*. This situation may occur at intervals as close as 19, 46, 65, 76, 122 and 141 years.

**Relation between Siderial and Tropical calendars (*ayanāṃśa*)**

It has been mentioned previously that the tropical or *nirayaṇa* year uses a dynamic initial point on the ecliptic, whereas the sidereal or *sāyana* year uses a fixed one. Over the years the distance between them increases. The Indian calendar calls this difference *ayanāṃśa*. This is the angular distance between the current equinox point and the fixed point of the Indian sidereal calendar.

Calculation of the *ayanāṃśa*

---

Western Astronomy is based upon the *sāyana* system, according to which the first point of the zodiac is defined to be the Vernal Equinox point. Now this point keeps on shifting relative to the fixed point of the sidereal zodiac. So the *aaynāṃśa* also keeps on increasing at the rate of about 50 1/3 arc-seconds every year. Thus, if we subtract this *ayanāṃśa* from the positions of planets and other asterisms given by the western system then we will arrive at the Indian positions.

When exactly the first points of the two zodiacs were chosen to coincide in India is not known exactly. Accordingly the *ayanāṃśa* or the precessional distance varies from 19° to 23°.

---

The following rules are given for conversion between the two systems.

(1) This method uses CE.

- Subtract 397 from the year in CE under consideration

- Multiply the remainder by 50 1/3 seconds; and reduce the product into degrees, arc-minutes and arc-seconds.
- Subtract this number of degrees, minutes, and seconds from the similar values given in the western system and the figure thus obtained will be according to the Hindu system.

Example: Determine the *ayanāṃśa* for 1912 CE.?

1912 - 397 = 1515 * 50 1/3 " = 76255"

76255" = 21° 10′ 55"

(2) This method uses *śaka* era (according to *grahalāghava*): subtract 444 from *śaka* era year, divide by 60, and express in angular units.

### *4.5 Traditional Lunisolar Year (vaiṣṇava)*

The *vaiṣṇava*s in Bengal belonging to the *gauḍīya vaiṣṇava* sect (founded by Lord *chaitanya*) of Hinduism use this calendar. This is almost the same as the North Indian lunisolar calendar in its structure. The nontraditional names of the calendrical elements remind one of the similarities to the Baha'I calendar.

The lunar year (*chāndra varṣa*) covers the period from *kṛṣṇa pratipadā* after the *gaura pūrṇimā* to the next *gaura pūrṇimā*. So this is a *pūrṇimānta* calendar. The fortnights are renamed *pradyumna* (the dark fortnight) and *aniruddha* (the light fortnight). The sun is named respectively as *balabhadra* when the sun is in its northern course or *uttarāyaṇa*, and *kṛṣṇa* when it is in its *dakshiṇāyaṇa* or southern course.

The months are named after the many names of Lord *viṣṇu*.

1/. *viṣṇu*, (*chaitra*, March-April)
2/. *Madhusūdana*, (*vaiśākha*, April-May)
3/. *trivikrama*, (*jyeṣtha*, May-June)
4/. *vāmana*, (*āṣāḍ ha*, June-July)
5/. *Śrīdhara*, (*śrāvaṇa*, July-August)
6/. *Hṛṣīkeśa*, (*bhādrapad*, August-September)
7/. *padmanābha*, (*āśvina*, September-October)
8/. *dāmodara*, (*kārttika*, October-November)
9/. *keśava*, (*agrahāyaṇa*, November-December)
10/. *Nārāyaṇa*, (*pauṣa*, December-January)
11/. *mādhava*, (*māgha*, January-February) and
12/. *govinda*, (*phālguna*, February-March).

Th *adhika māsa* (here called *puruṣottama māsa*) is inserted every two years and eight months. This additional month starts from the *śukla pratipadā* to the *amāvasya*.

The days of the week are also given various names of Lord *viṣṇu*:
1/ *sarva vāsudeva* (Sunday),
2/ *sarvaśiva sankarṣaṇa* (Monday),
3/ *sthāṇu pradyumna* (Tuesday),
4/ *yuta aniruddha* (Wednesday),
5/ *ādi kāraṇodakaśāyī* (Thursday),
6/ *nidhi garbhoadakaśāyī* (Friday),
7/ *abhaya kṣīrodakaśāyī* (Saturday).

The names of the *tithi*s according to the names of Lord *viṣṇu* are as follows;
1/ *pratipadā - brahmā*,
2/ *dvitīyā - śrīpati*,
3/ *tritīyā - viṣṇu*,
4/ *chaturthī - kapila*,
5/ *pancamī – Śrīdhara*,
6/ *ṣaṣṭhī - prabhu*,
7/ *saptamī - dāmodara*,
8/ *aṣtamī – hṛṣīkeśa*,
9/ *navamī - govinda*,
10/ *daśamī – madhusūdana*,
11/ *ekādaśī - Yudhar*,
12/ *dvādaśī - gadī*,
13/ *trayodaśī - śankhī*,

14/ *chaturaśī – padmī*,

15/ *pūrṇimā* and *amāvasyā - chakrī*.

# 5. The *panchān˙ga* or Indian Almanac

The almanac based on the principles of *jyotiṣa* is known as *panchān˙ga* in Sanskrit. This slim volume guides the religious and ceremonial life of Hindus everywhere in the world. The word *panchān˙ga* literally means "five limbs". These limbs are known as *vāra, tithi*, *nakṣatra*, *karaṇa*, and *yoga*. Their calculation follows astronomical principles but their significance is given by astrology. The *panchān˙ga* data is primarily used for determination of the religious festivals or fasting days. In addition, it is also used for horoscopes and for finding auspicious occasions for all kinds of activities like marriages, etc.

## *5.1 Initial Data, Time Units, and Geocentric View*

The time-related data is given in the sexagesimal units. This extends the unit of 60 beyond the familiar minutes.

1 *vipala* = 0.4 seconds

60 *vipala* = 1 *pala*=24 seconds

60 *pala* = 1 *ghaṭī* = 24 minutes

(*2 ghaṭī* = 1 *muhūrta* = 48 minutes)

60 *ghaṭī* = 1 *ahorātra* (= day-night) = 24 hours

## *5.2 The five basic elements of* **panchāṅga**

### (1) The *vāra* (The Day)

According to the scriptures of Hinduism, *brahmā* (the manifestation of *brahman* as Creator) started the creation of Universe at the sunrise and also the first day was a Sunday. The derivation of the other weekdays follows a rule. One assigns each *ghaṭī* (duration 24 minutes) to the next planet according to their geocentric order (Moon, Mercury, Venus, Sun, Mars, Jupiter and Saturn). This assignment is continued and the planet for the 61st *ghaṭī* names the next day. For instance, 1$^{st}$ *ghaṭī* of the starting day belongs to the Sun, next to the Mars, and so on until the 61$^{st}$ is assigned to the Moon and so the next day is Monday. Interestingly, if heliocentric order (Sun, Mercury, Venus, Earth+Moon, Mars, Jupiter and Saturn) is applied, the order of the days in the week using this method will be Sunday, Tuesday, Wednesday, etc.

The *panchāṅga* gives information about the name of the day and the time of sunrise and sunset (in hour and minute). It also gives the *dinamāna* or the difference between the sunset and sunrise (in *daṇḍa* and *pala*). The times of the sunrise and sunset are the same for a given longitude. They vary with the latitude and day of the year, which is given by the Equation of Time formula.

Surprisingly, not much significance is attached to the concept of a week that has become so important in the modern times. This has been taken to imply that the idea of a week is imported from the Semitic cultures in which it has a religious sanction. There may be some truth in this because the times of rest from studies in ancient times were some particular *tithi*s and not Sundays.

---

**(2) The *tithi***

The concept of *tithi* is unique to the Indian calendars. It is defined as the time taken for the angular separation between the sun and the moon to change by $12^0$. In other words, it is $1/30^{th}$ of the mean synodic period or about 0.98435 days (23 h 37 m 28 s). The actual value of *tithi* varies from a minimum of about 22 hours to a maximum of about 26 hours. It is calculated from the geocentric longitudes of the Sun and the Moon, and so it begins and ends at the same time throughout the world.

The day in Indian calendar starts at sunrise. The *tithi* in effect at sunrise is called the *tithi* of that day. As the length of a *tithi* is unequal to that of a day, so the following special situations arise:

(i) A *tithi* begins after a sunrise and ends before the next sunrise. In other words there is no sunrise during that *tithi*. Such a *tithi* is called a *kṣaya tithi*. In *panchāṅga*, such a *tithi* is omitted. Another name for this is *tryahasparśa* or touching of three days (by a *tithi*).

(ii) A *tithi* starts before the sunrise and ends after the next sunrise. It means that there are two sunrises during that *tithi*. Such a *tithi* is called an *adhika tithi* and is counted twice in the *panchāṅga*.

The *panchāṅga* gives the name of the *tithi*, its extent (in *daṇḍa* and *pala*) and the ending point of *tithi* (in hour and minute).

### (3) The *Nakṣatra*

The lunar orbit is inclined by about $5^0$ with the ecliptic. The passage of the moon around the ecliptic is divided into 27 parts (sometimes 28), each known as *nakṣatra*. The actual duration of a *nakṣatra* varies due to the complicated interaction of the sun, moon and earth. The names of *nakṣatra* with applicable month names in brackets are:

**The *Nakṣatra* Names**

| No. | Names | No. | Names | No. | Names |
|---|---|---|---|---|---|
| 1 | *Aśhvinī* (*Āśhvina*) | 10 | *Maghā* (*Māgha*) | 19 | *Mūla* |
| 2 | *Bharaṇī* | 11 | *Pūrvāphālgunī* | 20 | *Pūrvāṣāḍhā* (*Ā ṣā□□ha*) |
| 3 | *Krittikā* (*Kārttika*) | 12 | *Uttarāphālgunī* (*Phālguna*) | 21 | *Uttarāṣāḍhā* |
| 4 | *Rohiṇī* | 13 | *Hasta* | 22 | *Śravaṇā (Śrāvaṇa)* |
| 5 | *Mṛgśirā (Mārgaśīrṣa)* | 14 | *Chitrā (Chaitra)* | 23 | *Dhaniṣthā* |
| 6 | *Ardrā* | 15 | *Svātī* | 24 | *Śatabhiṣā* |
| 7 | *Punarvasu* | 16 | *Viśākhā (Vaiśākha)* | 25 | *Pūrvābhādrapada (Bhādrapada)* |
| 8 | *Puṣya (Ppauṣa)* | 17 | *Anurādhā* | 26 | *Uttarābhādrapada* |
| 9 | *Aśleṣā* | 18 | *Jyeṣthā (Jyeṣtha)* | 27 | *Revatī* |

Sometimes *abhijit* is also included as 28th in the sequence. The *panchān˙ga* gives the name and time period (in *daṇḍa* and *pala*) of *nakṣatra* in which the Moon can be seen that day. It also gives the ending time (in hour and minute) after which the Moon will move to the next *nakṣatra*.

**(4) The *Karaṇa***

This element is the half of a *tithi* so it is the time duration in which the angular difference between the longitudes of Sun and Moon is 6 degrees.

The first four are known as *sthira* or fixed *karaṇa*. There names and applicable tithi s are: *śakuni* (the 2nd half of *kṛṣṇa chaturdaśī*), *nāga*(the first half of *amāvasyā*), *chatuṣpada*(the second half of *amāvasyā*), and *kiṃtughna*(the first half of *śukla pratipadā*).

They are followed by the eight repeating cycles of the seven normal ones: *bava*, *bālava*, *kaulava*, *taittila*, *gara*, *vaṇij*, and *viṣti*. In this way all 60 half-*tithi*s in a month are covered.

They have basically astrological purpose only.

### (5) The *yoga*

This is taken to be the time duration in which the sum (or yoga) of the *nirayaṇa* longitude values of the Sun and the Moon becomes $13^{0}20'$. They are 27 in number. They have astrological significance.

**The Yoga Names**

| **No.** | **Names** | **No.** | **Names** | **No.** | **Names** |
|---|---|---|---|---|---|
| 1 | *Viṣkumbha* | 10 | *Gaṇ□□a* | 19 | *Parigha* |
| 2 | *Prīti* | 11 | *Vṛddhi* | 20 | *Śiva* |
| 3 | *Āyuṣmāna* | 12 | *Dhruva* | 21 | *Siddha* |
| 4 | *Saubhāgya* | 13 | *Vyāghāta* | 22 | *Sādhya* |
| 5 | *Śobhana* | 14 | *Harṣaṇa* | 23 | *Śubha* |
| 6 | *Atigaṇ□□a* | 15 | *Vajra* | 24 | *Śukla* |
| 7 | *Sukarmā* | 16 | *Siddhi* | 25 | *Brahmā* |
| 8 | *Dhṛti* | 17 | *Vyatīpāta* | 26 | *Indra* |
| 9 | *Śūla* | 18 | *Varīyān* | 27 | *Vaidhṛti* |

## *5.3* Other information in *Panchān˙ga*

### (1) *Ascendant or* lagna

The *lagna* or the Ascendant is the point of the ecliptic attached to the eastern horizon. This information is used by the astrologers.

For example, the *panchāṅga* shows that the *lagna* is 10 degrees of *karkạta rāśi*. This information is applicable with respect to local sidereal time as it depends on the local latitude. In sidereal system the *rāśi* divisions are fixed in the sky. This means that the point of ecliptic at 100 degrees from the initial point is intersecting the eastern horizon.

---

**(2)** ***Month and year***

The *panchāṅga* covers the year according to the prevalent era in that region of India. For instance the *Mithilā panchāṅga* covers one saala. The year starts at *śrāvaṇa kṛṣṇa pratipadā* (usually in July) as it follows the *pūrṇimānta* system. Additionally its era is equivalent to *Lakṣmaṇa* era of Bengal. In other parts of India the year unit may follow *Vikrama* era, and so on. The basic information as outlined above is similar (corresponding to the locally chosen latitude and longitude) but regional variations make them look very different. Still the beginning of other pan-Indian eras like *Vikrama*, *Śaka*, *Kali*, etc. are noted.

---

**(3)** ***The* rāśi *for the Sun and Moon***

The *panchāṅga* also contains the zodiacal position (*rāśi*) of the Sun at the time of sunrise (in *daṇḍa*, *pala*, and *vipala*). The name of *rāśi* is indicated numerically, e.g. 1 for *meṣa*, 2 for *vṛṣa*, etc. This information is called *sūryāṃśa* (part of the Sun).

As the Moon is seen moving very fast compared to the Sun, the lunar *rāśi* information is presented differently. The *rāśi* name is given explicitly and ending time is also given (in *daṇḍa* and *pala*).

---

**(4) *The* rāśi *for other planets***

The *rāśi* locations and coordinates for other planets (*Man˙gala*, *Budha*, *Guru*, *Śukra*, *Śani*) are also given sometimes. An interesting feature is the treatment of the lunar nodes (intersection of the orbits of the Earth and the Moon) as planets. They are known as *rāhu* and *ketu* and are very important for astrology. Their *rāśi* loacations and coordinates are also given.

---

### *5.4 Transformations for Other Locations*

The elements of the *panchān˙ga* vary in general with respect to locations.

- *vāra*: It depends on the local sunrise time.
- *tithi*: Its duration and count are global or same across the world. The starting and ending times are local.

- Other elements (*nakṣatra*, *karaṇa*, *yoga*) are treated in the same way as *tithi*.

### *5.5* Status of the "Week"

One very important observation is the omission of the week from the Vedic system. The word *saptāha* (or seven days) is a compound and does not occur in any of the astronomical considerations relevant in a *panchān˙ga*. It usually occurs in a religious context, for example *bhāgavata saptāha* (undertaking to recite *Śrīmad-bhāgavata mahāpurāṇa* in a week). In ancient India Sunday was not the day of rest. At the most, it was a very important day for the *saura* (devotees of the Sun-god as supreme) sect of Hinduism. The 5 ritual days of rest in a month are known as *pañchaka* and they are always mentioned in the *panchān˙ga*. Certain activities are forbidden during this period. The *anadhyāya* or no-study days for students were tied to *tithi*s and not to a weekday like Sunday.

One may conclude that the concept of the week and importance of Sunday as a rest day have been adopted by Hindus from non-Vedic cultures like Christianity and Judaism.

## 6. Different Eras and New Years of India

Indians have used various eras in their long history. Most of them were started by some king to commemorate his victory or coronation. They were used for the duration of that king's dynasty and were then replaced by the next ruler's era. They also had limited regional prevalence but a few of them attained pan-Indian acceptance. Currently three such eras are still being used by *panchāṅga* makers: *Vikrama*, *Śaka*, and *Kali*.

In this section some widely and also some lesser-known eras are described. The list is not at all comprehensive. This profusion of eras also highlights the lack of consensus among Indians about the need and choice of a common era for secular purposes.

## *6.1 Astronomical eras*

The astronomical eras have theological significance as their origin lies in the Vedic Cosmogony. Like every other religious tradition the Vedic account of Creation is unique. The era based on it is not used much in day-to-day life, but is implied in *saṃkalpa* (intention to undertake) of every Vedic rite like *pūjā*, *upanayana*, etc.

The basic repeating unit of the Creation in Vedic Cosmology lasts for 4 *parārdha*s, where one *parārdha* equals half of *Brahmā*'s life. The Creation takes place in two stages.

(i) The *sarga* or Primary Creation occurs at the beginning of the first *parārdha*. The *Brahman* (viewed as *Mahāviṣṇu, Paramaśiva,* and other manifestations by theists) starts the creation process by bringing out the 24 basic constituents (or *tattva*s) of both material and psychological cosmos (according to the *sāṃkhya* philosophy) into existence. These are: 5 *mahābhūta, 5 tanmātrā, 5 jñānendriya, 5 karmendriya,*

(ii) The *tattva*s are combined with *jīva*s and their *karma*s of the previous Creation. The Creator *Brahmā* springs forth and reorganizes the *sarga*-material into 14 layers of existence with their inhabitants in one particular *Brahmāṇ□□a*. This is known as *visarga* or Secondary Creation. This lasts for one *Brahmā* - day at the end of which a partial dissolution (*naimittika pralaya*) takes place. This is followed by one *Brahmā* – night. This sequence repeats for two *parārdha*s or for 100 *Brahmā* – years.

(iii) The total dissolution (*prākṛta* or *Brāhma pralaya*) takes place at the end of *Brahmā* life. The Universe is dissolved into its basic constituents and they themselves enter *Brahman*. This quiescent state lasts for two *parārdha*s at the end of which the cycle repeats again.

This Creation is quite different from the same term as understood by other religious traditions (e.g. Christianity, Islam, etc.). It is not creation out of nothing and it also does not start with the creation of elements (e.g. Light in Christianity) or the Earth. The values of many time periods associated with this process are given in the Appendix C.

Based on the above considerations the following three astronomical eras have been used by Indian astronomers: (i) *kaliyuga* era (ii) *manvantara* era, and (iii) *Sṛṣti* era.

### 6.1.1 The *kaliyuga* era = 5,105.

Out of these the *kaliyuga* era can be related to the traditional history as given in *purāṇa*s. Beginning of this era is associated with the following events:

- Events of the *Mahābhārata* war.
- Coronation of King *Yudhiṣthira* to the throne.
- Coronation of King *Parīkṣita* after 36 years.
- Disappearance of Lord *Kṛṣṇa.*

Among those scholars who accept that *Mahābhārata* war really took place in antiquity, there are genuine differences of opinion regarding its beginning. Currently there is a dearth of data so one can not reach an objective conclusion. Basically there are five different dates, which have been proposed.

- 5561 BCE. Date according to the astronomical calculations of P. V. Vartak.
- 3137 BCE. Date according to *Āryabhaṭa* and Aihole inscription.
- 2449 BCE. Date according to *Varāhamihira* with the traditional value to the *Śālivāhana śaka* era. Some historians claim that the *śaka* era mentioned by him (based on which the above date has been derived) refers to an earlier king. In that case this is not different from the date given by *Āryabhaṭa*.

- 1924 or 1424 BCE. Date according to Puranic genealogy with a guessed value for the gap between the War and the reign of Nandas.
- Around 1000 BCE. Date proposed according to the Aryan Invasion Theory proponents starting with Maxmueller and given in textbooks. Most Western historians agree with this date.

There is also a significant group of skeptical historians who do not accept its historicity at all. Whatever the final outcome of this debate, the *panchān˙ga* accepts the tradition and assigns a value to the *kaliyuga* era as first calculated by *Āryabhaṭa*.

According to this tradition, the *kaliyuga* era started on the 6$^{th}$ day after the departure of Lord *Kṛṣṇa*. The famous astronomer *Āryabhaṭa* noted its value in his time and currently that value is accepted for calendrical purposes. The *kaliyuga* itself started on *bhādrapada kṛṣṇa trayodaśī*, 20 February 3102 BCE, 2 hours 27 minutes 30 seconds. But that is not taken as the traditional starting point of the era itself. According to the accepted tradition, the 52$^{nd}$ century of *kaliyugābda* started in 1999 CE. On April 15$^{th}$, 2005, we are in the 5106$^{th}$ year of the current *kaliyuga*, which is only a small part of the current *chaturyuga* (4,320,000 years). The *kali* era is sidereal solar.

### 6.1.2 The *manvantara* era = 120,533,105.

The names of 14 *manvantara*s are given below:

- *PAST(6): Svāyambhuva*, *Svārochiṣa*, *Uttama*, *Tāmasa*, *Raivata*, and *Chākṣuṣa,*
- *CURRENT: Vaivasvat (7$^{th}$)*
- *FUTURE(7): Sāvarṇi*, *Dakṣa- sāvarṇi, Brahma- sāvarṇi, Dharma- sāvarṇi, Rudra- sāvarṇi, Deva- sāvarṇi,*and *Indra-sāvarṇi.*

One *Manvantara* period equals 308,448,000 years. This is equivalent to 71 *Mahāyuga*s plus a *sandhyā* of *kṛtayuga.* All of this equals 714 *kaliyuga*s. Currently 27 *Mahāyuga*s and 3 *yuga*s (of the 28$^{th}$ *mahāyuga*) have elapsed in the current seventh *Manvantara* named *Vaivasvata*.

#### 6.1.3 The *kalpa* or *Sṛṣti* (Creation) era = 1,971,221,106.

One *kalpa* is divided into 14 *Manvantara*s with additional years known as *kalpa sandhyā*. This can be more easily calculated in the units of *kaliyuga*s (1 *kaliyuga* = 432,000 siderial solar years). One *kalpa* lasts for a timespan of 10,000 *kaliyuga*s (or 4.32 billion years). It includes 14 *Manvantara*s (equaling 9996 *kaliyuga*s) and one *kalpa sandhyā* of one *kṛtayuga* or 4 *kaliyuga*s. Right now 4,563 (= 714 × 6 + 10 × 27 + 9) *kaliyuga*s have elapsed and we are currently in the 4,564$^{th}$ one. Putting it all together the *kalpa* era is found to be 1,971,221,106 (= 4,563 × 432,000 + 5,106) years. It can be favorably compared with the currently accepted scientific value of 13 to 14 billion years.

A still larger era is possible by adding *Brahmā*'s half-life to the above and getting 1,557,171,221,106 (= 1,555,200,000,000 + 1,972,949,106) years.

This is the oldest of all possible eras according to the Vedic cosmology. Like other two eras given earlier, it is rooted partly in the religious belief.

### 6.1.4 The *Bṛhaspati-chakra* or Cycles of Jupiter era.

Jupiter has a siderial period (with respect to fixed stars) of 11 years, 314 days, and 839 minutes, which is almost about 12 years. Its synodic period brings it in conjunction with the Sun every 398 days and 88 minutes, which is a little more than a year. Thus, Jupiter in about 12 years and Sun in one year pass through the same series of *nakṣatra*s. So a given year can be dated as the month of a 12 year cycle of Jupiter. A collection of 5 such cycles or 60 years is called *Bṛhaspati-chakra*, and each of the 60 years has distinct names. According to different traditions, it started in 427 CE or 3,116 BCE. Before 907 CE, one year was periodically omitted to keep the cycle in synchronization with solar years. Since 907 CE, the present year of the almanac is simply given the name of the year in the cycle.

Names of the years in a 60-year cycle are given below.

| No. | Name | No. | Name | No. | Name |
|---|---|---|---|---|---|
| 1 | *Prabhava* | 21 | *Sarvajit* | 41 | *Plavan˙ga* |
| 2 | *Vibhava* | 22 | *Sarvadhāri* | 42 | *Kīlaka* |
| 3 | *Śukla* | 23 | *Virodhī* | 43 | *Saumya* |
| 4 | *Pramudita* | 24 | *Vikṛti* | 44 | *Sādhāraṇa* |
| 5 | *Prajotpatti* | 25 | *Khara* | 45 | *Virodhakṛt* |
| 6 | *An˙girasa* | 26 | *nandana* | 46 | *Paridhāvī* |
| 7 | *Śrīmukha* | 27 | *Vijaya* | 47 | *Pramādhīśa* |
| 8 | *Bhava* | 28 | *Jaya* | 48 | *Ānanda* |
| 9 | *Yyuvā* | 29 | *Manmatha* | 49 | *Rākṣasa* |
| 10 | *Dhātu* | 30 | *Durmukha* | 50 | *Nala* |
| 11 | *Īśvara* | 31 | *Hemalam ī* | 51 | *Pin˙gala* |
| 12 | *Bahudhānya* | 32 | *Vilambī* | 52 | *Kālayukti* |
| 13 | *Pramati* | 33 | *Vikā ī* | 53 | *Siddhārthī* |
| 14 | *Vikrama* | 34 | *Śarvarī* | 54 | *Raudrī* |
| 15 | *Viṣu* | 35 | *Plava* | 55 | *Durmukhī* |
| 16 | *Chitrabhānu* | 36 | *Śubhakṛt* | 56 | *Dundubhi* |
| 17 | *Svarbhānu* | 37 | *Śobhakṛt* | 57 | *Rudhirodgārī* |
| 18 | *Taraṇa* | 38 | *KrodhI* | 58 | *Raktākṣī* |
| 19 | *Pārthiva* | 39 | *ViŚvāvasu* | 59 | *Krodhana* |
| 20 | *Vyaya* | 40 | *Parābhava* | 60 | *Kṣaya* |

The *panchā˙nga* gives the name of the current year according to the above table. This era is appropriate for the span of an average human life. The Chinese also follow a similar Cycle of Jupiter era. It is hard to say in which direction the borrowing took place. It may also have been independently discovered by both India and China.

### 6.1.5 The *Saptarṣi* or *laukika* era.

It is based on the assumption that the *saptarṣi* or Ursa Major (Big Bear) stars complete one revolution around the North Pole star in 100 years. This is astronomically approximate only, but was adopted anyway. It is supposed to have started in 3076 BCE or 25 years after the beginning of the *kali* era. A given century in this era starts on the 76th year of the corresponding Christian century.

This era has been mentioned by Alberuni in his memoirs of India. It is still prevalent in Kashmir and Punjab.

## *6.2 Historical Eras*

These eras are named after famous kings or religious figures. They are still followed  in different regions of India by one or other group. Two of these (*Vikrama* and *Śaka* eras) have pan-Indian acceptance.

### 6.2.1 Buddhist or *Bauddha* Era

This era is named after Lord Buddha, who is one of the greatest figures of world history. Despite all efforts, the dates of the Buddha's birth and death remain uncertain. Three dates are available with respect to King Ashoka's coronation (273 BCE).

- The Sri Lankan sources place the Buddha's birth in 623 BEC, and his death in 543 BCE, but these dates are rejected by most Western and Indian historians.
- The dates attested by all Sanskrit and Chinese sources, places Buddha's death 100 years before Ashoka's consecration.
- Western historians place Buddha's death 210 years before Ashoka's consecration, so according to them He was born in 563 BCE and died in 463 BCE.

Buddhist concepts of great temporal periods is similar in spirit to Hinduism but differs in details. Originally the largest unit was called *kalpa* of 4,320,000 years (equivalent to a *mahāAyuga* in Vedic terms). Later its name was changed to *mahākalpa*. It lasts 1,344,000,000 (1.344 Gyr) years. A *mahākalpa* is divided into 4 regular *kalpa*s (also called *asaṃkhyeya kalpa*), which contain formation, existence, destruction, and emptiness of the world systems. It lasts 336,000,000 or 336 million years. Each of the four such *kalpa*s is divided into 20 *antarā-kalpa*s one of which lasts 16,800,000 or 16.8 million years.

### 6.2.2 Jain or *Vīrasamvat* Era

This era started in 528 BCE after the nirvana of Lord *Mahāvīra*, who is the 24th *Tīrthan˙kara* of Jain religion and also the senior contemporary of Lord Buddha. The year begins on *Dīpāvalī*. This calendar is followed by the *Jaina* community.

The Jain cosmology is very different from the Vedic one and is not based on astronomical considerations. The universe is considered to be eternal and uncreated. The largest temporal unit is called *kalpa* but it has a different meaning in this system. A *kalpa* is divided into two equal periods called *utsarpiṇī* (ascending phase) and *avasarpiṇī* (descending phase). Each one of them is further divided into three *arā*s (arcs of time), which last for vast periods of time. The timespan of an *arā* has not been given precisely. The six *arā*s are:

(iv) *Suśama*- *Suśama*: the phase of absolute happiness.

(v) *Suśama*: the phase of happiness

(vi) *Suśama –Duḥśama* the phase of much happiness and some sorrow

(vii) *Duḥśama* - *Suśama*: the phase of much sorrow and some happiness

(viii) *Duḥśama*: the phase of sorrow

(ix) *Duḥśama* - *Duḥśama* the phase of absolute sorrow.

The larger unit of *kalpa* keeps on repeating eternally. According to Jain beliefs, right now we are in the 5th phase of the present *kalpa*.

### 6.2.3 Vikram Era or *samvat*

This is the most well-known era used currently in traditional India. It ha many versions:

(i) In the north India, it begins with *chaitra*, and each month begins with the full moon (*pūrṇimā*).

(ii) In Gujarat, the year begins with *kārttika* and month begins with the new moon (*amāvāsyā*).

(iii) In some parts of Gujarat, the year begins with the new moon of *Āṣāḍha*.

There is some controversy about the identity of the king who promulgated this era. Most popular belief credits *Chandragupta,* the king of Ujjain, who took the title of *Vikramāditya* to have started this lunisolar era in 57 BCE. He is not to be confused with the more famous king of the Gupta dynasty with the same name.

### 6.2.4 *Śālivāhana Śaka* Era

Other important era is *Śaka* or *Śālivāhana*, which can be either solar or lunisolar. In the north it is reckoned as elapsed and in south as current. The lunisolar months of this era begin with full moon (*pūrṇimā*) in the north and with the new moon (*amāvasyā*) in the south. There is also some controversy regarding who started it. It was thought that the *Kuṣāṇa* king *Kaniṣka* started this era in 78 CE, but based on new findings some historians now argue against it.

The Indian National *Śaka* era: This era has been adopted by the Government of India. Its elapsed year is tropical solar and it begins on the day following the vernal equinox. The first month is *chaitra*, with 30 days in a normal year and 31 in a leap year. The next 5 months have 31 days and the rest have 30 days.

### 6.2.5 Kalachuri Era

This era commemorates a ruler of South Indian Kalachuri (or *Kālachūri*) dynasty. It started in 248 CE and is not in vogue now.

### 6.2.6 Gupta Era

The golden age of Guptas is commemorated by this era. It started in 321 CE, but it is not in vogue now.

### 6.2.7 Bengali San and other related Eras

This era was supposed to have started in 593 CE in the reign of the king of Bengal (called *Gouda* at that time) named *Śaśān˙ka.* There is some controversy about whether he started it himself or it was coincidental. It is possible that the king *Bhāskaravarman* of *Prāgjyotiṣapura* ascended to the throne in that year and started the *Bhāskarābda* era still used in Assam; and this may have influenced the Bengali choice. Later its structure was changed during the reign of Mughal king Akbar under the influence of the calendar taareekh-e-Ilaahi promulgated by him in 1584 CE.

Though that calendar died out soon, the corresponding Bengali calendar (also known as fasli or crop-related) maintained the same era. It is based on the model of the planetary motions given in *Sūrya- siddhānta*, and the number of days in the months vary from year to year.

Some other eras, which came into existence around the same time and follow more or less the same system, are given below.

- Amli (this starts on *Bhādra śukla dvādaśī*, in Orissa),
- Vilayati (592 CE),
- Fasli (590, 592, or 593 CE, in Bihar, this starts on *Bhādra kṛṣṇa pratipadā*),
- Sursan (599 CE, in Maharashtra).

- *Bhāskarābda* (same as San, Assam),
- *Tripurābda* (in Tripura)

### 6.2.8 Harṣa Era

This era is named after the last emperor of ancient India (before Islamic invasion), the King Harshavardhan (or *Harṣa-vardhana*) of Kannauj. It started in 647 CE. It is not used now.

### 6.2.9 Kollam Era

It started in 825 CE, and is connected with Lord Parashuram (or *Paraśu-rāma*) and is prevalent in Malabar Coast and Tirunelveli. Its years are current and solar. They start with entry of Sun in *kanyā* (or Virgo) in north Malabar or in *siṃha* (or Leo) in south Malabar. It is sometimes divided into cycles of 1,000 years reckoned from 1176 BCE. Thus 825 CE is the first year of the 3rd Millennium. This era is widely used in *panchāṅga*s from Kerala.

### 6.2.10 Nepali Era

Nepal follows the Vikram Samvat (known as Bikram Samvat or BS in Nepal) in its sidereal solar form and is similar to the Bengali calendar. The year starts around April 14th and so the first month is *Vaiśākha*. A parallel lunar calendar is used for determining the religious dates.

# 7. Epilog

All Indian calendars follow a basic template for reconciling a sidereal solar year with a synodic lunar year. The underlying mathematical models are the same or very similar. The apparent differences are superficial as they are related to the differences in month names, month beginnings, etc. The main problem facing us is to base them on correct observations and redefine their starting points so that the discrepancy with seasons is removed. One of the Appendixes describes one such proposal.

Next task is to convince Indians to adopt it. The task is easy on paper but very difficult in practice. People become very comfortable with existing calendars and it is almost impossible to convince them to follow a changed one for a dimly perceived ideal of logical and mathematical consistency. At the same time it must be remembered that progress always implies a break with status quo and also sometimes painful adjustment with the change. It is hoped that many more people will reach similar conclusions about the need to reform Indian calendars and when a critical mass will be built, it will be very easy to implement the reform.

# Appendix A: Astronomical Constants

Many different definitions of a month are given in the table below.

| Name | Definition | Present Value |
|---|---|---|
| Synodic Lunar Month (LM) | Time between the one full Moon and the next as seen from Earth (also called one lunation). | 29d 12h 44m 3s (29.5308d) = 29.53058815 d |
| Siderial Lunar Month (SM) | Moon's period of revolution around Earth as seen in the FSF. | 27d 7h 43m 11s (27.32166d) |
| Tropical Month (TM) | Time between successive passages of the moon through the $1^{st}$ point of the Aries. | 27.32158 d |
| Anomalistic Month | The interval between successive perigees, which are the configurations when the distance between moon and earth is the shortest. | 27.55455 d |
| Draconic month | Period of revolution of the moon through the nodes (the points of intersection of moon's plane of rotation with the ecliptic) of its orbit. | 27.21222 d |
| Lunar Year (LY) | Length of 12 mean lunations | 354.24219 d (it is 10.8751 days shorter than tropical year) |
| Siderial Year (SY) | The period after which the Earth's axis returns precisely to the same position in the FSF. | 365.25636 d = 365d 6h 9m 10s. Mean number of lunations in an SY = ? |
| Tropical Year (TY) | Time between successive passages of the Sun through the $1^{st}$ point of the Aries as seen on Earth (= time between start of seasons in consecutive years) | 365.2421875 d = 365 d 5 h 48 m 45.00 s. Mean number of lunations in a TY = 12.36827. (TY length varies by about 20 minutes from year to year due to nutation and planetary interactions). |
| Anomalistic year | The interval between successive perihelions, which are the configurations when the distance between Sun and earth is the shortest. | 365.25964 d |

It should be noted that the FSF values for day and month are larger than the Earth ones, but for a year it is the other way around. The reason for this is the phenomena of the Precession of the Equinoxes.

Different values were given for the length of the Siderial Year (SY) different *siddhānta* treatises, as shown in the table below.

**Table 3-3:**

| **siddhAnta** | **Length of the SY (d)** | **Number of years after which SY = TY + 1 d** |
|---|---|---|
| Aryabhaṭīya | 365.258681 | 61 |
| Rājamṛgānka | 365.258691 | 61 |
| Sūrya-siddhānta | 365.258756 | 60 |

---

# Appendix B: Divisions of Time in Ancient India

- System I according to Jain Mathematicians

| | |
|---|---|
| 256 *Avalikā* | = 1 *Kṣullaka bhava* |
| 65,535 *Kṣullaka bhava* | = 3,773 *Prāṇa*<br>= 16,777,216 *Āvalikā* |
| 3,773 *Prāṇa* | = 1 *Muhūrta* = 48 minutes<br>= 0.8 hr |
| 30 *Muhūrta* | = 1 day = 24 hours |

- System II according to Jainism

| | |
|---|---|
| 1 *Paramāṇu-kāla* (pk) | = 16/18,225 sec |
| 1 *Dvyaṇu- kāla* (dk) | = 2 pk = 32/18,225 sec |
| 1 *Tryaṇu- kāla* (tk) | = 3 dk == 32/6,075 sec |
| 1 *Truṭi- kāla* (trk) | = 3 tk = 32/2,025 sec |
| 1 *Vedha- kāla* (vk) | = 3 trk = 32/675 sec |
| 1 *Lava- kāla* (lk) | = 3 vk = 32/225 sec |
| 1 *Nimeṣa- kāla* (nk) | = 3 lk = 32/75 sec |
| 1 *Kṣaṇa- kāla* (kk) | = 3 nk = 32/25 sec = 1.28 sec |
| 1 *Kāṣthā- kāla* (Kk) | = 5 kk = 6.4 sec |
| 1 *Laghu- kāla* (lgk) | = 15 Kk = 1.6 minutes = 96 sec |
| 1 *Ghaṭī* (g) | = 15 lgk = 24 minutes = 0.4 hr |
| 1 *Muhūrta* (m) | = 2 *Ghaṭī* = 48 minutes = 0.8 hr |
| 30 *Muhūrta* | = 60 *Ghaṭī* = 1 day (= 24 hours) |

- System I according to *Bhāskarāchārya*

| | |
|---|---|
| 1 *Truṭi* | = 1/33750 sec |
| 100 *Truṭi* | = 1 *Nimeṣa* (n) |
| 1 *Asu* or *Prāṇa* | = 45 n = 4 sec |
| 18 n | = 1 *Kāṣthā* (k) |
| 30 k | = 1 *Kalā* (kl) |
| 30 kl | = 1 *Ghaṭī* (g) |

- System II according to *Bhāskarāchārya*

| | |
|---|---|
| 6 *Prāṇa* | = 1 *Vināḍī* |
| 6 *Vināḍī* | = 1 *Nāḍī* (= ghaTI) |
| 2 *Nāḍī* | = 1 *Muhūrta* = 48 minutes |

- System used in *panchā˙nga*

| | |
|---|---|
| 60 *Vipala* | = 1 *Pala* |
| 60 *Pala* | = 1 *Ghaṭī* |
| | = 1 *Daṇḍa* |

## Appendix C: Cosmic Cycles in Vedic Cosmology

In this Appendix, the infrastructure of Cosmic Cycles as envisioned in the scriptures of Hinduism is explained by tables.

**(i) Structure of a *Mahāyuga***

| | |
|---|---|
| #days (number used in *Sūrya-siddhānta*) | 1, 577, 917, 800 |
| #solar months | 51, 840, 000 |
| #solar years = #revolutions of Sun | 4, 320, 000 |
| #days / #solar years | 365.25875 |
| #revolutions of moon (or #lunar months) | 57, 753, 336 |
| #lunar revolutions (or months) - #solar revolutions (or years) | 53, 436, 336 |
| #tithis | 53, 436, 336 x 30 = 1, 603, 000, 000 |
| #tithis/#days | 703/692 |

**(ii) One *Chaturyuga* in divine (DY) and Siderial Solar Years (SSY)**

| Yuga | Dawn | Duration | Twilight | Total |
|---|---|---|---|---|
| *Kṛta* | 400 DY = 144,000 SSY | 4,000 DY = 1,440,000 SSY | 400 DY = 144,000 SSY | 4,800 DY = 1,728,000 SSY |
| *Tretā* | 300 DY = 108,000 SSY | 3,000 DY = 1,080,000 SSY | 300 DY = 108,000 SSY | 3,600 DY = 1,296,000 SSY |
| *Dvāpara* | 200 DY = 72,000 SSY | 2,000 = DY = 720,000 SSY | 200 DY = 72,000 SSY | 2,400 DY = 864,000 SSY |
| *Kali* | 100 DY = 36,000 SSY | 1,000 = DY = 360,000 SSY | 100 DY = 36,000 SSY | 1,200 = 432,000 SSY |
| Total | 1,000 DY = 360,000 SSY | 10,000 = DY = 3,600,000 SSY | 1,000 DY = 360,000 SSY | 12,000 DY = 4,320,000 SSY |

**(iii) Structure of a *Manvantara* with *Sandhyā***

| **periods** | **#in a *Manvantara*** | ***Sandhyā* (Twilight)** | **Total** |
|---|---|---|---|
| *Chaturyuga*s | 71 | 0.4 | 71.4 |
| *Kaliyuga*s | 710 | 4 | 714 |
| SSY | 306,720,000 | 1,728,000 | 308,448,000 |

**(iv) Structure of *Manvantara*s in a *Kalpa***

| | |
|---|---|
| 1 *Manvantara* | 71 *Mahāyuga* + 1,850,000 SSY = 308,570,000 SSY |
| *Pralayakāla* | 2,000 yrs + 12 * 1,500 yrs = 20,000 yrs = 13 inter-*Manvantara pralaya*- occurrences |
| 14 *Manvantara*s with *Pralaya-kāla* | 4,320,000,000 yrs = 4.32 Gyr = 1 *kalpa* |

**(v) Structure of a *Kalpa***

| **periods** | **#in 14 *Manvantara*s** | **Introductory dawn** | **Total** |
|---|---|---|---|
| *Chaturyuga*s | 999.6 (=71.4*14) | 0.4 (=1 *Kṛtayuga*) | 1,000 |
| *Kaliyuga*s | 9996 | 4 | 10,000 |
| SSY | 4,318,272,000 | 1,728,000 | 4.32 Gyr |

**(vi) *Brahmā*'s life**

| | |
|---|---|
| 1 *Brahmā* day (= 1 *Brahmā* night = 1 *kalpa*) | 1,000 *Mahāyuga* = 4,320,000,000 SSY (or 4.32 Gyr) |
| 1 *Ahorātra* of *Brahmā* | 8.64 Gyr |
| 1 year of *Brahmā* (=360 *ahorātra* of *brahmA*) | 360*8.64 Gyr = 3,110.4 Gyr = 3.1104 Tyr |
| One *parārdha* (= 50 *Brahmā* yrs) | 155.52 Tyr |
| *Brahmā*'s lifespan (or 100 yrs) | 311.04 Tyr |
| The basic repeating unit of Cyclic Creation | 622.04 Tyr |

**(vii) Right now on 14th April, 2009**

| | |
|---|---|
| **First half (*Pūrvārdha*) of *Brahmā*'s life past** | |
| 50 years | = 1,555,200,000,000 years (or 1,555.2 Gyr) |
| **1st day of 1st year of 2nd half (*Parārdha*) of *Brahmā*'s life or 13 *ghaṭī* 42 *pala* past.** | |
| Six *Manvantara*s (*Svāyambhuva*, *Svārochiṣa*, *Uttama*, *Tāmasa*, *Raivata*, and *Chākṣuṣa*) past | = 6 × 308,448,000 yrs = 1,850,688,000 yrs (or 1.850668 Gyr) (i) |
| In 7th (*Vaivasvata*), 27 *Mahāyuga*s are past = 27 *X* 4,320,000 yrs | = 116,640,000 yrs (ii) |
| In 28th *Mahā-yuga* 3 *yuga*s (*kṛta*, *tretā*, *dvāpara*) past | = 9 × 432,000 yrs = 3,888,000 yrs (iii) |
| Total number of sidereal solar years past before the beginning of the present *kaliyuga* (28th) | = (1) + (ii) + (iii) = 1,971,216,000 yrs |
| Total number of sidereal solar years past in this *kalpa* on 14th April 2009 | = (1) + (ii) + (iii) + 5,106 (year-count of *Kali*-era) = 1,971,221,110 yrs |
| Yrs of the 7th *Manvantara* past on | = (ii) + (iii) + 5106 = 120,533,110 |
| 14th April 2009 | |
| Yrs of the 7th *Manvantara* remaining | = 188,036,890 |
| Yrs from the beginning of this *Śveta-vārāha Kalpa* past on 14th April 2009 | = 1,955,885,110 (year-count of *Sṛṣti* era) or 1.955885110 Gyr |

Right now we are in the first day of the 51st year of *Brahmā* 's life and some time of that day has also elapsed. The 52nd century of sidereal solar *kali*-era started on 14th April, 1999 CE. According to tradition, the *kaliyuga* started at 20 February 3102 BCE, 2 hours 27 minutes 30 seconds. Total number of sidereal solar years past in this current *Śhveta-vārāha kalpa* (White Boar *kalpa*) on 14th April 2009 = 1,971,221,110 yrs.

---

## Appendix D: Precession of Equinoxes and Vedic Cosmic Cycles

A very interesting relationship between the length of the tropical year, and the value of the precession of the equinoxes has been found by Dwight W. Johnson. It is also tied to the Vedic Cosmology at the same time. The following discussion has been taken from his website.

*Relationship between SY and TY*

One should note that the equinoxes (the points of intersection of ecliptic and equator) move by about 50 Seconds in one year (= 360/25,800 degrees) or about 30 degrees in 2,150 years. Starting point for this calculation is the value of the Constant of Precession (or the amount of precession in degrees in one SY) taken as 7/500 degrees (same as $0.014^0$ or 50.4”). This is used to find some precession-related values given in the tables below.

| Precessional Periods | Duration |
|---|---|
| 1 Precessional Day<br>= time period for $1^0$ of precession | 500/7 SY (= 71 + 3/7) SY |
| 1 Precessional Year (PrY)<br>= time period for $360^0$ of precession | 180,000/7 (= 25,714 + 2/7) SY |

Similarly the relation between some Siderial time-periods and their precessional equivalents is given in the next table.

| Time Periods | Value of Precession-periods |
|---|---|
| 1 SY | 7/500 degrees (Constant of Precession) |
| 1 *chaturyuga* (or 4,320,000 SY) | 4,320,000 × 180,000/7 = 168 PrY |
| 1 *manvantara* (or 71 3/7 *chaturyuga*) | 168 × 500/7 = 12,000 PrY |
| 1 *kalpa* (= 14 *manvantara*) | 14 × 12,000 = 168,000 PrY |

In one *chaturyuga*, there are 4,320,000 SY. The number of TY in the same time period can be obtained by adding the 168 PrY to this number. The total number of days is the same so we get the following relation.

*365.2563795 (days/SY) × 4,320,000 SY = (#days/TY) × 4,320,168 TY*

So the number of days in a TY

= 4,320,000 × 365.2563795)/4,320,168 = 365.2421756.

This value is quite near to the currently accepted one of 365.242199 days and better than that used by the Indian *panchā˙nga* –makers.

*Comparison of Precessional Constants*

| **Entity** | **Hindu (H)** | **Newcomb (N)** | **Difference (H-N)** | ***Sūrya-siddhānta (s)*** | **Difference (H-s)** |
|---|---|---|---|---|---|
| Prec.Const. | 50".4/SY | 50".2583/SY | 0".0417/SY | 54"/SY | 3".6/SY |
| SY | 365.2563795 | 365.2563627 | 1".4 /SY | 365.258756 | -205".4 /SY |
| TY | 365.2421756 | 365.2421988 | -2".0 /SY | | |

As we can see, modern Indian astrology uses a very inferior set of values.

## Appendix E: Saros Cycle and Eclipses

An eclipse occurs when the cycles of the following two events coincide.

- Moon is at one of the two nodes determined by the Draconic month.
- Moon is aligned with the Earth and the Sun determined by the Synodic month.

It is found that

- 223 Synodic months = 6585.321 days.
- 242 Draconic months = 6585.357 days

This period, of 18 years, 10 and a third days, is known as the **Saros**; during which similar eclipses (ie. eclipses belonging to the same Saros cycle) re-occur.

There is slight difference of about one-third of a day between the two periods. The Earth rotates this extra one-third between the two Saros cycles. This results in the Saros cycle eclipses taking place over different parts of the Earth. This also means that the eclipses will repeat over a larger period of three times the Saros period (called Triple Saros), which equals 54 years and 32 or 33 days (depending on leap years).

Finally, there are many different combinations of circumstances which can cause an eclipse: Moon at its rising node, Moon at the falling node, etc.; and this is why there are 42 Saros cycles running at any one time, and hence why we get several eclipses (solar and lunar) per year.

## Appendix F: Equation of Time

We know that the time of sunrise and sunset varies with latitude and day of the year. Equation of Time (EOT) is the mathematical formulation of this phenomenon.

EOT = Sundial time - Clock time = True sun time – Mean sun time

The time as obtained from a watch is the "Mean Sun Time" based on the average of solar motion in a year. The time obtained from the true solar motion is different from this. To get a clearer picture of this phenomenon, let us look at the watch-time when the Sun crosses the Prime Meridian at a given place. One would guess that the time interval between such crossings on successive days will be exactly 24 hours but it is not so. Actually this varies during the year from being ahead by about 17 minutes to being behind by about 14 minutes.

These differences result due to the tilt of the Earth's rotation axis (23.5 deg) and the eccentricity of its elliptical orbit around the Sun. It can be seen as an asymmetric figure of eight on globes and is obtained from the sum of offset sine-functions. EOT is needed for astrological calculations based on *panchā˙nga*.

## Appendix G: Comparison of SSY with ABPSS proposal

Proposal by the Akhil Bharatiya Panchanga Sudhara Samiti (ABPSS) has the following elements and they are contrasted with the current proposal in the table given below.

| Topic | ABPSS proposal | Our proposal |
|---|---|---|
| Model vs. observation | The calendrical calculations will use the most recent data from observational astronomy | Agrees. |
| Solar year | Tropical | Sidereal. This is the most important difference between the two proposals. |
| Connection with VJ | The *uttArayaNa* (around Dec 21) is identified with the beginning of the solar *mAgha* (or Vedic *tapah*) and also that of *Śiśira Rtu*. Then the *vasanta Rtu* and solar *chaitra* (or Vedic *madhu*) has to start from | Agrees. |
| | around Feb 20. | |
| Beginning of the lunar year | The Lunar New Year begins on the 1st *amAvasyA* after the start of solar *chaitra* (or Vedic *madhu*). | Agrees. |
| Type of lunar month | The *amAnta* and *pUrNimAnta* traditions of lunar month in different parts of India will be maintained. | There should be only one type of lunar month and that should be *amAnta*. |
| Beginning of the solar year | The Vernal Equinox (*vasanta-sampAta*) around March 21 is the beginning of solar *vaiśAkha* (or Vedic *mAdhava*) and also that of the Solar New Year. | Agrees. |
| Status of *rAśi*s and | The *rAśi*s are deemed to be foreign imports into the Vedic dharma so they are not accommodated. | The *rAśi*s are accommodated due to the sidereal nature of the proposed calendar. Connection between the solar month and beginning of *rAśi*s will be broken so April 15 can still be the beginning of *meSHa* but not the beginning of any solar month. Or if desired, *meSHa* can be redefined to start with the solar new year thus breaking the connection between *rAśi* and *nakṣatra*. |
| Attitude to astrology | Very much opposed to astrology. | Neutral to astrology. This proposal will accommodate it if so desired. |
| Status of *nakṣatra*s | Connection to *nakṣatra*s will be maintained in some modified form. Most probably their beginning and duration will be changed. | The traditional equal length divisions will be maintained, but the beginning may be changed. |
| Periodic | There is no such need as the year is already tropical. | One day every 71 year to be removed from the SY to |
| Synchronization | | synchronize it with the TY. |

This proposal will break the following correspondences being followed in the current *panchAnga*s.

(i) The connection between the lunar month and ideally the rule of *pUrNimA* being in the middle of *nakṣatra* after which the lunar month is named, e.g. the new lunar *chaitra* will not contain *chitrA nakṣatra*. Even currently the *pUrNimA* of a given lunar month does not fall in the middle of the related *nakṣatra*. This deviation took place because the sidereal year and lunar year are connected by the rules of synchronization so the deviation in the solar year is reflected in the lunar months. The table below shows the current situation.

| Lunar month | Current position of moon on *pUrNimA* | Lunar month | Current position of moon on *pUrNimA* |
|---|---|---|---|
| *chaitra* | | *āśvina* | |
| *vaiśākha* | | *kārttika* | |
| *jyeṣ.tha* | | *mārgaśIrṣa* | |
| *āṣāḍha* | | *pauṣa* | |
| *śrāvaNa* | | *māgha* | |
| *bhādrapada* | | *phālguna* | |

(ii) The Connection between *rāśi* and *nakṣatra* will be broken if the former are redefined to start with the SSY.

(iii) The connection between the sidereal *rāśi* and beginning of solar months, e.g. the beginning of *mādhava* will not coincide with the beginning of the *meSha*.

## Appendix H: A Proposal for a new Hindu Calendar

After surveying the different Indian Calendars, one wishes that like other national groups Hindu-Indians should also have one unifying calendar. A proposal in this regard is presented below for consideration.

**Why reform?**

Suppose nothing is done and we live with the status quo regarding the calendar. Why should it be a problem?

The first problem is that discrepancy between the seasons and the calendar will widen with the passage of time. Some cultures and traditions are not bothered by it. The Islamic lunar calendar rotates throughout the solar year and the Orthodox Christianity is still following the Julian calendar. But Vedic Hindu tradition insists on the correspondence between seasons and religious days. Some examples follow:

(a) Makara Sankranti is supposed to denote the beginning of *uttarāyaNa*, but right now it is off by about 3 weeks.

(b) The lunar year is supposed to start with the spring in the Northern Hemisphere. Right now in Northern India it starts when it already feels like summer.

The second problem is the fact that the current astronomical knowledge is not used in the calculations of Hindu calendars with a few exceptions. The methods being used may have been the best when they were discovered but then they were not updated with the passing times. This was mostly due to the loss of political power by Hindus.

The third problem is the bewildering variations on the same basic calendar templates. They obstruct any attempt to unify the Hindu society because these different calendars divide the Hindu society at a very basic level.

**Some religious considerations:**

The *panchān˙ga* is not a simple compilation of astronomical data, but it is such data in a context. At the bottom, this context is theological. In Vedic thought, the universe itself has aspects of divinity. The different schools of *vedānta* characterize this relation between Brahman and the Cosmos differently. A majority of them (*advaita*, *viśiṣṭādvaita*, *dvaitādvaita*, *achintya-bhedābheda*, etc.) do not accept the reality of the world separate from that of the *Brahman*. Only *dvaita* school posits them differently. But all philosophical schools agree that divinity has expressed itself through time and space and its modality was intuited by the *Ṛṣi*s or seers of *veda*s. It is they who decreed that special angular configuration of the Sun, the Moon, the planets and the *nakṣatra*s have meanings beyond mere secular and astronomical. It is on their authority expressed through the scriptures that tradition believes that *ekādaśī* is holy to Lord *Viṣṇu*, or *saṃkrānti* is appropriate for special *pūjā*s, etc.

Many great saints and reformers like Kabir, Nanak and others have said that there is nothing holy about these temporal configurations. A little deeper investigation reveals that many of them view the creation as ontologically different from the Creator. As Time is a created entity, so it can not contain any divine attributes. Thus they vociferously rejected this aspect of the received wisdom and declared that any time or place is equally good or bad for any activity. So they could not see God in a stone or holiness in a river. This view is in tune with semitic religions like Christianity and Islam. The mainstream did not agree with this logic and carried on unchanged. Any proposal for a new *panchā ̇nga* has to respect these religious and cultural considerations.

Some recommendations follow for a proposed new Indian calendar.

***Recommendation* 1: The solar year should be Siderial**

Apart from being Lunisolar, the year also should be Sidereal, so that the *rāśi*s and *nakṣatra*s are defined from a fixed point in the sky. The tropical zodiac starts from the spring equinox point, which moves at a constant rate in the sky with respect to the sidereal fixed point. The adoption of the tropical system will create a break from the past and will go against the religious sentiments of the current tradition. Probably this is the reason behind the rejection of the national calendar by the masses.

Some people are advocating the new reformed calendar to be tropical because the Vedic corpus talks about tropical months like *madhu*, *mādhava*, etc and also about the tropical years starting from solstices and/or equinoxes. It should be noted that as Hinduism developed this approach was abandoned very early and there was an India-wide consensus to base the calendars on sidereal principles.

The biggest drawback of uncorrected SY is its loss of connection with the seasons. But now as this slippage is correctly understood, the corrections can be made to the SY to maintain its correspondence with the seasons. Even the TY needs such periodic corrections but at larger time intervals.

***Recommendation* 2: The first year of the proposed Siderial Solar Year (SSY) should begin at the Spring Equinox**

The tropical year is defined by two solstices and two equinoxes. They are based on observable astronomical phenomena. In principle the SSY can start from any one of them. The Vedic tradition talks about either the Winter Solstice or the Spring Equinox as the start of the year. It is proposed that the new SSY should start on the day of Spring Equinox as this will be the least disruptive.

***Recommendation* 3: The proposed SSY should be synchronized with the tropical year and thus with the seasons about every 71 years**

As the sidereal year (SY) is a little longer than the tropical year (TY), this difference amounts to one day in approximately 71 years, which is not going to be significant in an average person's lifetime. This difference can not be predicted accurately over very long time periods because there is a stochastic component to it. But if it is neglected, then over centuries, the SY will not be synchronized with the seasons determined by the TY any more. As an example, the current discrepancy of three weeks between the start of seasons between the two will become about five weeks within next 1000 years.

So there should be a rule to reconcile the SY with TY by dropping one day or a few days from the sidereal year at regular intervals. One practical rule can be the following: The year, in which the difference (SY-TY) at the beginning of the TY (taken here as the vernal equinox) is more than 24 hours, SY will start one day earlier at the correct vernal equinox point. The *ayanāṃśa* (the angular distance on the ecliptic between the spring equinox point and the fixed sidereal point) during these 71 years will be less than one arc-minute. The calendar reform is rooted in astronomical considerations but one has to be cognizant of the need of the astrology as it is a very important part of the Indian scene. They will have the freedom to either adopt the new sidereal fixed point according to the rules indicated above or live with the old one.

The fixed point of the SY has been moved around in the past and now it is a good time to do that again. Currently it is such that it was coincident with Vernal Equinox in 285 AD. This date has no historical or astronomical significance and is arrived at by back calculation so that it leads to the current *ayanāṃśa* used by astrologers.

The first *saṃkrānti* of the current year will be at the vernal equinox point. The next point after $30^0$ with respect to the fixed star background will be the second *saṃkrānti*, and so on. This redefinition will bring the seasons in synch with the newly defined SY.

There will be a break in the definition of years if these rules are adopted. The year in which the reform is made will be an abnormal one because many days will be removed from it. This undertaking will be similar to the Gregorian reform of the Common Era calendar. It is a small price to pay for a unified calendar of the Vedic tradition. We can call this system of counting the year as Synchronized Siderial Year (SSY). It accepts the fact that we have to synch up with the tropical year at regular intervals (by losing 1 day every 71 year and re-aligning the initial point of the zodiac to VE) to have correct seasons.

The proposal given above results in loss of one day once about every 71 years due to forced synchronization of the sidereal and tropical years. There may be an objection that this will disrupt the continuous counting of the *nirayaṇa* year sequence. The year losing a day due to synchronization will not be a *nirayaṇa* but will become almost a *sāyana* year except that it will begin at the end of the previous sidereal year.

So the year count will have the 70 sidereal years followed by an anomalous year that will be neither tropical nor sidereal and then the cycle will start all over again. This is a sticking point in this scheme as both orthodox and modern astronomers will dislike this mixed up counting rule. But then one can point out that there is nothing sacrosanct about the older year count and it can be replaced by the sequence proposed above.

There may be an objection that the year count coming from the start of the *kaliyuga* will be interrupted. But the idea that the count has been continuous and started from the same beginning is a myth. It should be pointed out that even the Vedic records mention differing starting points of years. Also the traditional count itself does not include the dynamic effects discovered in the last few centuries like the cycles involved with the orbital eccentricity and the orbit position with respect to the invariant plane of the solar angular momentum (these are the effects due to the Milankovic ice age cycles). These effects modify the length of the SY and so the traditional count is already inaccurate. Based on these considerations, the modified count of era should be acceptable.

***Recommendation* 4: The sidereal solar months will be renamed to avoid confusion with existing month names.**

The same names for both lunar and solar months are used these days leading to much confusion. This can be removed by using Vedic names for sidereal solar months and common names for lunar months as given in the table below.

| **Solar month (new names)** | **Season (*ṛtu*) (Northern Hemisphere)** | **Begins around** | **Comments** |
|---|---|---|---|
| *madhu* | *vasanta* | Feb 21 | The Lunar New Year begins on the 1st *amāvasyā* falling between starts of solar *madhu* (Feb 21) and solar *mādhava* (March 21) |
| *mādhava* | " | March 21 | Beginning of the sidereal solar year equated with that of Vedic month *mādhava*.<br><br>The beginning of first year of the the proposed new calendar is exactly on *vasanta-sampāta* (Vernal Equinox).<br><br>Beginnings of the years coming after the first one will drift at the rate of precession until it is reset to the VE in the 72nd year. |
| *śuchi* | *grīṣma* | April 21 | |
| *śukla* | " | May 21 | |
| *nabha* | *varṣā* | June 21 | *dakṣiṇāyana* (Summer Solstice) |
| *nabhasya* | " | July 21 | |
| *iṣa* | *śarada* | August 21 | |
| *ūrja* | " | September 21 | *Śarada-sampāta* (Autumnal Equinox) |
| *saha* | *hemanta* | October 21 | Winter begins |
| *sahasya* | " | November 21 | |
| *tapah* | *śiśira* | December 21 | Equivalnce of solar *māgha* (old name) and Vedic month *tapah* based on VJ. Also the beginning of *tapah* equated to *uttarāyaṇa* (Winter Solstice). |
| *tapasya* | " | January 21 | |

***Recommendation* 5: The duration of the solar months in the proposed SSY should be determined according to rules given below**

Earlier the solar months were defined as the duration between two consecutive solar *saṃkrānti*s with respect to *rāśi*s. In SSY system we have the following choices:

*1st Choice*: The duration of solar months should be tied to the months of the Gregorian or common calendar. Then one of the months will have one extra day in a leap year. Same criticism as of the first choice applies here.

*2nd Choice*: The four quarters of the year based on the cardinal points are not equal. Each quarter should be divided into three equal months as much as possible and any extra day can be added to the last month. The time taken by the Sun to move from one cardinal point of the year to the another, and proposed duration of months can be chosen as follows.

(i) Spring Equinox to Summer Solstice = 92.8d. There will be 3 months of 31 days each.

(ii) Summer Solstice to Autumn Equinox = 93.6 d. There will be 3 months of 31 days each.

(iii) Autumn Equinox to Winter Solstice = 89.8 d. There will be 3 months of 30 days each.

(iv) Winter Solstice to Spring Equinox = 89.0 d. There will be 2 months of 30 days and 1 month of 29 days. During leap years all 3 months will have 30 days each.

*3rd Choice*: The solar month will be the duration between the two points on the ecliptic $30^0$ apart starting at the beginning of the year. As the Earth's orbital speed varies during the year, so the duration of solar months will vary and will not equal those of the Gregorian calendar. They will also not overlap with the solar months based on the equal division sidereal zodiac. At one time it was difficult to calculate this but now astronomical observations are available to determine them directly.

The last two choices are almost equivalent. The 3rd choice will be adopted as it is easier to calculate and least disruptive to the present definitions of sidereal solar month.

The names of the new solar months will be taken from Vedanga Jyotisha and are given in Appendix H. Originally they were used for the tropical solar months but we will use the same names for the sidereal solar months. According to our rules they will almost always overlap and the difference will be at the most of one day.

***Recommendation* 6: The lunar year connected to the proposed SSY should begin**

**(i)** **after the first** *amāvasyā*

**(ii)** **between the** *madhu saṃkrānti* **(currently same as** *sāyana mīna saṃkrānti* **occurring around February 20), and the** *mādhava saṃkrānti* **(around Vernal Equinox or March 21)**

This rule is obtained after identifying the tropical solar *māgha* or the start of the *śiśira ṛtu* with the winter solstice point usually around Dec 21. This rule was followed during the era of Vedanga Jyotish of *Lagadha* and will be used here. Beginning of the spring season (*vasanta ṛtu*) then falls around February 20. If the lunar year has to begin in the spring then it must start after the first *amāvasyā* after the February *saṃkrānti* (same as *madhu saṃkrānti*).

It is to be noted that the *saṃkrānti*s of SSY are not connected with the zodiac names.

***Recommendation* 7: The lunar months will be** *amānta* **and the current rules for determining** *adhimāsa* **and** *kṣayamāsa* **will be applied to them in conjunction with SSY**

A choice must be made for the definition of a lunar month acceptable by Vedic Hindus all over the world. One can find quotes from the *śruti* and *smṛti* to justify both the *amānta* and the *pūrṇimānta* systems of the lunar month. But now is the time to choose one system and make it the standard for all Hindu calendars. It is proposed, that the *amānta* system should be chosen because the most prevalent *Vikrama* era starts with the beginning of a month in this system.

The lunisolar year is tied to the SY with very elegant rules, so there is no need to change them. Same rules will be applied to new lunar months in conjunction with the SSY.

### ***Recommendation*** **8: The lunar months will be renamed**

The lunar months have to be renamed after its alignment with SSY. The traditional names like *caitra, vaiśākha,* etc. are connected to the *nakṣatra*s. The *caitra pūrṇimā* for example has to fall within *citrā nakṣatra* as a guiding principle and so on with other lunar months. Now new lunar months of SSY are not connected to *nakṣatra*s. So their names should be different to avoid any confusion. The following names originating from ancillary Vedic scripture are proposed.

| **Lunar month (new names)** | **Start** | **Comments** |
|---|---|---|
| *aruṇa* | After around Feb 21<br>(In the proposed calendar, Lunar New Year will precede the Siderial | The Lunar New Year begins on the 1st *amAvasyA* after the start of solar *madhu* i.e. after around Feb 21 |
| | New Year) | |
| *aruṇarāja* | Later lunar months will follow the succeeding *amAvasyA*s. These new lunar months will not be anchored to *rāśi*s or *nakṣatra*s. | 2nd Lunar Month |
| *puṇDarIka* | | 3rd Lunar Month |
| *viśvajit* | | 4th Lunar Month |
| *abhijit* | | 5th Lunar Month |
| *ārdra* | | 6th Lunar Month |
| *pinvamāna* | | 7th Lunar Month |
| *unnvamāna* | | 8th Lunar Month |
| *rasavāna* | | 9th Lunar Month |
| *irāvAna* | | 10th Lunar Month |
| *sarvauShāDha* | | 11th Lunar Month |
| *sambhara* | | 12th Lunar Month |

These names have also been used in a calendar published by Dr. Ravi Prakasha Arya. Originally these names were alternative set in the later Vedic times and were used as solar month names. In the proposed scheme I have taken the liberty of using them for lunar months.

***Recommendation* 9: The *rāśi* will not be used in the proposed calendar**

The concept of *rāśi* is not needed for the proposed calendar's smooth functioning. The rules of *dharma-śāstra* do not require them as they were written before the *rāśi*s were accepted by Indian astronomers during there interaction with foreign systems. Those rules are concerned with *nakṣatra*s mostly. But the Indian tradition of calendar making has carried the baggage of *rāśi* for the past 2000 years and it is very difficult to disentangle them. The astrologers need them and they have become ingrained in the popular Hinduism.

The proposed calendar gives the astrologers two options.

- First one will be the easiest and it consists in identifying the proposed sidereal solar months with the *rāśi* names. These will follow the *sāyana* system and will be practically indistinguishable from Western one.
- The second one is to identify the correct period of solar months coincident with the current *nirayaṇa rāśi* divisions. There will be no complete overlap of these solar months with any such *rāśi*.

***Recommendation* 10: All 27 *nakṣatra*s will have equal angular division on the ecliptic.**

The concept of *nakṣatra*s is unique to Indian calendars. In the ancient times they were tied to prominent stars along the ecliptic and so had unequal angular extensions. Later the divisions were made equal and tied to *rāśi*s with one *rāśi* being equal to 2¼ *nakṣatra*s. They are connected to important holy days and festivals.

These *nakṣatra* divisions many times do not contain the defining prominent stars after which they are named. Many purists would like to go back to unequal divisions and redefine their boundaries. It is doubtful if this exercise will be acceptable to the astrologer community. The unequal divisions were made into equal ones for the ease of calculations in the first place.

It is recommended that the current divisions should be accepted. Their relation to the proposed calendar will be through the *ayanāṃśa* which is the distance between VE and beginning of *aśvinī.*

| *Nakṣatra* Name | Theoretical equal-length extent | *Yogatārā* Name | Comments |
|---|---|---|---|
| *Aśhvinī* | $0^0$-$13^0$20’ | α or β Arietis | First in the current lists |
| *Bharaṇī* | $13^0$20’-$26^0$40’ | 41 Arietis | |
| *Krittikā* | $26^0$40’-$40^0$ | η Tauri (Alcyone) | |
| *Rohiṇī* | $40^0$-$53^0$20’ | α Tauri (Aldebaran) | |
| *Mṛgśirā* | $53^0$20’-$66^0$40’ | λ Orionis | |
| *ārdrā* | $66^0$40’-$80^0$ | α Orionis (Betelgeuse) | |
| *Punarvasu* | $80^0$-$93^0$20’ | β Gemin - orum | |
| *Puṣya* | $93^0$20’-$106^0$40’ | δ Cancri | |
| *āśleṣā* | $106^0$40’-$120^0$ | ε Hydrae | |
| *Maghā* | $120^0$-$133^0$20’ | ρ Leonis (Regulus) | |
| *Pūrvā-phālgunī* | $133^0$20’-$146^0$40’ | δ Leonis | |
| *Uttarā-phālgunī* | $146^0 40'$-$160^0$ | β Leonis (Denebola) | |
| *Hasta* | $160^0$-$173^0 20'$ | δ Corvi | |
| *Chitrā* | $173^0 20'$-$186^0 40'$ | α Virginis (Spica) | |
| *Svātī* | $186^0 40'$-$200^0$ | α Bootis | |
| *Viśākhā* | $200^0$-$213^0 20'$ | α Librae | |
| *Anurādhā* | $213^0 20'$-$226^0 40'$ | δ Scorpii | |
| *Jyeṣthā* | $226^0 40'$-$240^0$ | α Scorpii (Antares) | |
| *Mūla* | $240^0$-$253^0 20'$ | λ Scorpii | |
| *Pūrvāṣāḍhā* | $253^0 20'$-$266^0 40'$ | δ Sagittarii | |
| *Uttarāṣāḍhā* | $266^0 40'$-$280^0$ | σ Sagittarii | |
| *Śravaṇā* | $280^0$-$293^0 20'$ | α Aquilae (Altair) | |
| *Dhaniṣthā* | $293^0 20'$-$306^0 40'$ | β Delphini | |
| *Śatabhiṣā* | $306^0 40'$-$320^0$ | λ Aquarii | |
| *Pūrvā-bhādrapada* | $320^0$-$333^0 20'$ | β Pegasi | |
| *Uttarā-bhādrapada* | $333^0 20'$-$346^0 40'$ | γ Pegasi | |
| *Revatī* | $346^0 40'$-$360^0$ | ζ Piscium | |

***Recommendation* 11: The *kali* era should be used universally and others should be gradually de-emphasized.**

The regional variation in the era name should be allowed but there should be some universal era acceptable to all Hindus. It is proposed that *kali* era should be prominently used for all different *panchā˙nga* published in the country. In addition the *Sṛṣṭi* or Creation era can also be used as it is used by implication in all religious rites through *saṃkalpa*. This has one problem and that is its difference from the scientific creation era, which is supposed to have started about 14 Billion years ago. As long as both are not confused with one another it is acceptable. These eras are accepted universally throughout India and are sanctioned by the religious tradition. The *Vikrama* and *Śaka* eras do not capture the antiquity and continuity of the Vedic astronomical tradition so they should be gradually de-emphasized. Other regional eras should be also removed from the general usage over time.

***Recommendation* 12: The *panchā˙nga* makers all over India should be requested to use correct observational data and not the incorrect formulas from older *jyotiṣa* works.**

We should base our *panchā˙nga* on observations made at the national astronomical observatory or NASA/JPL. The previous year's data can be fitted to yield a simple mathematical expression, which can be used by the *panchā˙nga* -makers. The national center for Positional Astronomy should handle this task. This should be a huge improvement over the outdated mathematical models used at the current time. Because of the inaccuracies in these models, the published values of the start, duration and end of the *panchā˙nga* elements is wrong most of the time and this violates the spirit of the *jyotiṣa*.

Other elements of *panchā˙nga* like *vāra, tithi, kara ṇa,* and *yoga* will be unaffected by these considerations.

These recommendations are very modest and can be implemented without much disruption. They will synchronize the new calendar with the seasons and bring about a much-needed cultural unity to the Vedic fold.

## Appendix I: Mathematics behind the *panchā˙nga* Calculations

The basic five elements are calculated with varying degrees of difficulties.

The input parameters for calculations are the following:

(i) The time at which the values are needed

= $T_G$ (in unit of hours after previous midnight)

= GMT time corresponding to the local time T.

(ii) The GMT Siderial Solar Longitude (in degrees) at

a. midnight before $T_G = S_B$

b. midnight after $T_G = S_A$

These values are given in NASA/JPL's Astronomical Ephemeris.

(iii) The GMT Siderial Lunar Longitude (in degrees) at

a. midnight before $T = M_B$

b. midnight after $T = M_A$

These values are given in NASA/JPL's Astronomical Ephemeris.

(iv) Sunrise time at a given latitude $L = T_{SR}(L)$

Intermediate values assuming uniform solar and lunar motion

(i) The GMT time corresponding to $T = T_G$

(ii) The SSL and SLL at $T_G$

$S(T_G) = S_B + (S_A - S_B)*(T_G/24)$

$M(T_G) = M_B + (M_A - M_B)*(T_G/24)$

Vernal Equinox is the zero point of the ecliptic (a fixed point on ecliptic for sidereal calculation) from which longitudes are calculated

Calculation of *vāra* $v(T_G)$

$v(T_G) = S(T_G)*360/SY$

Here SY = length of Siderial Year = 365.256366 days

At a given latitude L, $v(T_G)$ will be given by the above if $T > T_{SR}(L)$

Calculation of *tithi* number (tn) from *tithi* $t(T_G)$,

$t(T_G)$ =Integer next to $|S(T_G) - M(T_G)|/12$

Alternately tithi = (((360 + M - S) mod 360)/12) + 1.

At a given latitude L, $t(T_G)$ will be given by the above if $T > T_{SR}(L)$

Calculation of *nakṣatra* $n(T_G)$

Let *ayanāṃśa* = A

Then $n(T_G)$ = Integer next to $[A+M(T_G)]*3/40$

Calculation of *kara ṇa* $k(T_G)$

This element is equal to half a tithi, so its calculation is tied to $t(T_G)$. Their total number is 60 with four fixed and 7 cycles of 8 varying values.

Calculation of *yoga* $y(T_G)$

$$y(T_G) = \text{Integer next to } [S(T_G)+M(T_G)]*3/40$$

They have 27 values.